\documentclass[letterpaper,11pt]{article} 

\usepackage{amsmath}
\usepackage{amssymb}
\usepackage[comma,longnamesfirst]{natbib}
\usepackage[dvips]{epsfig}
\usepackage{dcolumn}
\usepackage{enumerate}
\usepackage{hhline}
\usepackage{dsfont}
\usepackage{afterpage}
\usepackage{arydshln}
\usepackage{graphicx}
\usepackage{adjustbox}
\usepackage{color}
\usepackage[usenames,dvipsnames]{xcolor}
\usepackage{rotating}
\usepackage[breaklinks]{hyperref}
\usepackage{breakurl} 
\usepackage{xr}
\usepackage[percent]{overpic}
\usepackage{subfig}
\usepackage[]{caption}
\usepackage{algorithmicx}
\usepackage[noend]{algpseudocode}
\usepackage{algorithm}
\usepackage{diagbox}
\usepackage{graphicx}
\usepackage{wrapfig}
\usepackage{lscape}
\usepackage{fontenc}
\usepackage{setspace}
\usepackage{bm}
\usepackage{slashbox}
\usepackage{lscape}
\usepackage{breakurl} 
\usepackage{multirow,booktabs}
\usepackage{eurosym}
\usepackage{titlesec}
\usepackage{sectsty}
\usepackage{placeins}
\epsfverbosetrue
\def\theequation{\thesection.\arabic{equation}}  
\def\abstract{\if@twocolumn
\section*{Abstract}
\else \normalsize 
\begin{center}
{\bf Summary\vspace{-.5em}\vspace{0pt}} 
\end{center}
\quotation 
\fi}
\def\endabstract{\if@twocolumn\else\endquotation\fi}

\makeatletter
\newcommand{\myappendix}[1]{
	\setcounter{section}{1}
        \renewcommand{\thesection}{A\arabic{section}}}

\def \dvec {\text{\boldmath$d$}}

\def \uvec {\text{\boldmath$u$}}

\def \yvec {\text{\boldmath$y$}}    
\def \zvec {\text{\boldmath$z$}}

\def \gammavec        {\text{\boldmath$\gamma$}}
\def \deltavec        {\text{\boldmath$\delta$}}
\def \epsilonvec      {\text{\boldmath$\epsilon$}}
\def \varepsilonvec   {\text{\boldmath$\varepsilon$}}

\def \thetavec        {\text{\boldmath$\theta$}}

\def \lambdavec       {\text{\boldmath$\lambda$}}

\def \pivec           {\text{\boldmath$\pi$}}

\def \psivec          {\text{\boldmath$\psi$}}

\def \omegavec        {\text{\boldmath$\omega$}}

\usepackage{color}
\usepackage{colordvi}
\newlength{\breite}
\breite\textwidth
\newcounter{aufg}[section]
  {\refstepcounter{aufg}\noindent\textbf{Exercise \arabic{aufg}:}
   \\*[1ex]\noindent}{\vspace{.5cm}}
   
 \newcounter{notes}[section]
  {\refstepcounter{aufg}\noindent\textbf{}
   \\*[1ex]\noindent}{\vspace{.5cm}}
   
\usepackage{amsthm}  

\theoremstyle{definition}

\newtheorem*{beisp*}{Example}
\newtheorem{Proof}{Proof}


\newtheoremstyle{break}
  {}
  {}
  {}
  {}
  {\bfseries}
  {.}
  {\newline}
  {}
  
\theoremstyle{break}

\newcommand{\head}[2]%
 {\hrule \vspace{.15cm} {\sfbold Advanced Statistical Inference, Summer Term 2012, Georg-August-University G\"ottingen}\hfill
{\sfbold Sheet #1}\\
{\sfbold Prof. Dr. Thomas Kneib, Nadja Klein}\hfill {\sfbold #2}

\vspace{.2cm}
\hrule

\vspace{1cm}

}

\newcounter{auf}
{\refstepcounter{auf}
\begin{center}
\fcolorbox[gray]{0}{.95}{
\makebox[\breite]{
\textbf{Exercise \arabic{auf}}
}}\\*[1ex]\noindent
\end{center}
}{\vspace{.5cm}}

\newcounter{loes}[section]
{\stepcounter{loes}
\begin{center}
\fcolorbox[gray]{0}{.95}{
\makebox[\breite]{
\textbf{L"osung \arabic{loes}}
}}\\*[1ex]\noindent
\end{center}
}{}

{\begin{center}
\fcolorbox[gray]{0}{.95}{
\makebox[\breite]{
\textbf{Zu Aufgabe #1}
}}\\*[1ex]\noindent
\end{center}\vspace{1cm}
}{\vspace{1cm}}

\newcounter{ka}
{\refstepcounter{ka}
\begin{center}
\framebox[\textwidth]{
\textbf{Aufgabe \arabic{ka}} \hfill #1 Punkte
}\\*[1ex]\noindent
\end{center}
}{\vspace{1cm}}

\newcounter{lka}
{\refstepcounter{lka}
\begin{center}
\framebox[\textwidth]{
\textbf{L\"osung \arabic{lka}} \hfill #1 Punkte
}\\*[1ex]\noindent
\end{center}
}{\vspace{1cm}}

\usepackage[margin=1in]{geometry}
\titlespacing*\section{0pt}{0pt plus 4pt minus 2pt}{0pt plus 2pt minus 2pt}
\titlespacing*\subsection{0pt}{0pt plus 4pt minus 2pt}{0pt plus 2pt minus 2pt}
\titlespacing*\subsubsection{0pt}{0pt plus 4pt minus 2pt}{0pt plus 2pt minus 2pt}

\definecolor{myblue}{RGB}{0,73,114}
\allsectionsfont{\sffamily\color{myblue}}

\newcounter{myremark}

\newcounter{mynotation}

\usepackage{paralist}

\renewenvironment{itemize}[1]{\begin{compactitem}#1}{\end{compactitem}}

\makeatletter
\def\@seccntformat#1{\@ifundefined{#1@cntformat}%
	{\csname the#1\endcsname\quad}  
	{\csname #1@cntformat\endcsname}
}
\let\oldappendix\appendix 
\renewcommand\appendix{%
	\oldappendix
	\newcommand{\section@cntformat}{\appendixname~\thesection\quad}
}
\makeatother

\usepackage{titlesec}

\usepackage{scalerel,stackengine}
\stackMath
\newcommand\reallywidehat[1]{%
\savestack{\tmpbox}{\stretchto{%
  \scaleto{%
    \scalerel*[\widthof{\ensuremath{#1}}]{\kern-.6pt\bigwedge\kern-.6pt}%
    {\rule[-\textheight/2]{1ex}{\textheight}}
  }{\textheight}%
}{0.5ex}}%
\stackon[1pt]{#1}{\tmpbox}%
}

\begin{document}
\setlength{\abovedisplayskip}{0.15cm}
\setlength{\belowdisplayskip}{0.15cm}
\pagestyle{empty}
\begin{titlepage}

\title{\bfseries\sffamily\color{myblue}  
	Implicit copula variational inference}
\author{Michael Stanley Smith and Rub\'en Loaiza-Maya}
\date{\today}
\maketitle
\noindent
{\small Michael Smith is Professor of Management (Econometrics) in the Melbourne Business School, University of Melbourne, Australia. Rub\'en Loaiza-Maya is Senior Lecturer in the Department of 
	Econometrics and Business Statistics, Monash University, Australia. Correspondence should be directed to Michael Smith at {\tt mikes70au@gmail.com}.

\noindent \textbf{Acknowledgments:} We thank Professor David Nott for discussions on the use of spherical co-ordinates. We
also thank two referees, an Associate Editor and the Editor, Professor Galin Jones, for providing comments
that improved the manuscript. We are grateful to Professor Mark Frank for placing his inequality data in the public domain.}\\

\newpage
\begin{center}
\mbox{}\vspace{2cm}\\
{\LARGE \title{\bfseries\sffamily\color{myblue} Implicit copula variational inference}
}\\
\vspace{1cm}
{\Large Abstract}
\end{center}
\vspace{-1pt}
\onehalfspacing
\noindent
Key to effective generic, or ``black-box'', variational inference
is the selection of an approximation
to the target density that balances accuracy and speed.
Copula models are promising options, but calibration of the approximation 
can be slow for some choices. 
\cite{smith2020trans} suggest using tractable and scalable
``implicit copula'' models that are formed by element-wise transformation of the
target parameters. We propose an adjustment to these transformations
that make the approximation invariant to the scale and location of the target
density. We also show
how a sub-class of elliptical copulas have a generative representation that allows
easy application of the re-parameterization trick and efficient
first order optimization.
We demonstrate the estimation methodology using 
two statistical models as examples. 
The first is a mixed effects logistic regression, and the second is a regularized correlation matrix. For the latter, 
standard Markov chain Monte Carlo estimation methods can be slow or difficult to implement, yet our proposed variational approach provides an effective and scalable estimator. 
We illustrate by estimating a regularized Gaussian copula model for income inequality in U.S. states between 1917 and 2018. An Online Appendix
and MATLAB code to implement the method are available as Supplementary Materials. 
\vspace{20pt}
 
\noindent
{\bf Keywords}: Elliptical copula; Factor model approximation; Re-parameterization trick; Spherical co-ordinates; Stochastic gradient ascent; Variational Bayes. 
\end{titlepage}

\newpage
\pagestyle{plain}
\setcounter{equation}{0}
\renewcommand{\theequation}{\arabic{equation}}

\section{Introduction}\label{sec:intro}
Variational inference (VI) methods are increasingly
used for the Bayesian estimation
of big or complex statistical models.
They require the selection of a variational approximation (VA) to the posterior distribution
and an optimization approach for its calibration. Recent focus is 
on so-called ``black box'' VI methods~\citep{ranganath14} that
combine generic VAs and optimization methods that require little or no
tailoring to the statistical model being estimated. One promising black
box approach is ``copula VI''~\citep{tran2015,han+ldc16,smith2020trans,gunawan2021},
where 
copula models are used as generic VAs. Copula models are multivariate distributions 
constructed from marginals and a copula function; see~\cite{nelsen06} 
for an introduction. 
However, for many choices of copula function 
and/or marginals, the resulting VA can be computationally expensive to calibrate, limiting their adoption.
To address this problem, 
\cite{smith2020trans} show that ``implicit copula'' models
can be 
flexible, scalable and fast to calibrate. 
The objective of this paper is to refine, extend and demonstrate the efficacy of their approach.

An implicit copula is constructed from a known continuous multivariate 
distribution $\psivec\sim F$
by inverting Sklar's theorem; see~\cite{smith2021}. 
An attractive feature for VI is that
implicit copula models can be represented as a one-to-one element-wise transformation
between $\psivec$ and the model parameters $\thetavec$ (also
 called ``latent variables'' in the machine learning literature).
\cite{smith2020trans} show that selecting tractable parametric forms
for this transformation and the
distributional family of $F$,
allows ready application of the ``re-parameterization trick''. This trick
facilitates efficient implementation of stochastic gradient optimization algorithms for calibration
of the implicit copula VA. We
extend this idea here in three ways.
First, the transformations suggested by these authors are shown to produce
VAs that have a level of accuracy that varies
with the location and scale of $\thetavec$, which is 
a poor property. An adjustment to the transformations are proposed
to correct for this.
Second, for $F$ we consider the sub-class of elliptical distributions
that can be expressed as a scale mixture of normals.
These have a generative representation that is amenable to 
the re-parameterization trick. 
Third, we follow~\cite{miller2017,ong+ns16,mishkin2018slang} and others by adopting
a sparse factor (also called a ``low rank plus diagonal'') decomposition for the scale matrix $\Sigma$ of distribution $F$. 
To identify
the elliptical copula $\Sigma$ should have a leading diagonal of ones, and we show
how to use spherical co-ordinates to impose this bound.
The end result is a class of
elliptical copula model VAs that is scalable and fast to calibrate.


Other copula VI methods include that suggested by 
\cite{tran2015} who use vine copulas. Calibration in 
high dimensions is typically slow compared to implicit copulas. 
\cite{chi21} suggest speeding up Monte Carlo 
gradient estimates in vine copula VI using draws under a mean
field simplification. However, such estimates are more 
noisy than those computed using an effective re-parameterization gradient.
The simplest implicit copula is the Gaussian copula,
which was first employed as a VA by
\cite{han+ldc16}. \cite{smith2020trans} extend
this to the skew-normal copula and exploit sparse factorizations of the form
in~\cite{ong+ns16} for the 
scale matrix and the re-parameterization trick. 
\cite{gunawan2021} generalize the VI method of~\cite{smith2020trans} to a copula formed from a mixture of normals. Each mixture component is calibrated sequentially using a variational boosting algorithm, although the 
re-parameterization gradients can remain noisy in this case~\citep{miller2017}
and more computationally demanding estimates are necessary.
The transformations suggested in our paper can also 
be used to make the VA of~\cite{gunawan2021} invariant to changes in
the location and scale of $\thetavec$. \cite{hirt2019} propose a flexible VA that is motivated by
(but is not)
a copula model. As with our approach, their proposed VA has a 
generative representation that allows efficient implementation 
of stochastic gradient optimization.

To illustrate the increase in accuracy that our VA can provide 
compared to that in~\cite{smith2020trans}, we re-examine their
mixed logistic regression example. But our main application is the 
estimation of a regularized correlation 
matrix $\Omega$, which is an important problem in statistical
modeling~\citep{rothman08}. Following~\cite{rebonato1999}
and others, we re-parameterize 
$\Omega=LL^\top$ using spherical co-ordinates for the Cholesky factor $L$. We then 
then use a horseshoe prior~\citep{carvalho2010} for a transformation
of these angles
that provides adaptive shrinkage of 
partial correlations towards zero. Estimation of the joint posterior
of $\Omega$ and the shrinkage hyper-parameters using standard MCMC methods can be prohibitively slow. However, implicit copula VI can be employed
to successfully estimate $\Omega$ in higher dimensions than MCMC. 
We show how to do so for a regularized Gaussian copula correlation matrix for annual income inequality
in 49 U.S. states between 1917 and 2018. A rich pattern of interstate
dependence that mirrors geographical and socio-economic commonalities 
between states is observed. A t-copula model VA is shown to capture the posterior more accurately 
than either a t-distribution or mean field approximations.


The rest of the paper is organized as follows. Section~\ref{sec:cvi} gives
a brief introduction to copula VI, with a focus on implicit copula
VI. Section~\ref{sec:lVA} details our 
extension and refinement of the methodology of~\cite{smith2020trans}.
Section~\ref{sec:examples} contains the two examples, while Section~\ref{sec:discuss} concludes. Appendix~\ref{app:A} gives
the closed form gradients required to implement the SGA algorithm. An Online Appendix provides further details on notation used, the derivatives, examples, and MATLAB code.
\section{Copula variational inference}\label{sec:cvi}
\subsection{Variational inference}
Consider a statistical model with continuous parameters $\thetavec=(\theta_1,\ldots,\theta_m)^\top$, data $\yvec$, 
and  posterior density $p(\thetavec|\yvec)\propto p(\yvec|\thetavec)p(\thetavec)=g(\thetavec)$.
Variational inference approximates $p(\thetavec|\yvec)$ with 
a density 
$q(\thetavec)$ that is called a ``variational approximation'' (VA).
The approximation is obtained by minimizing a divergence measure between $q$ and the posterior, with the Kullback-Leibler (KL) divergence the most common choice.
In this case,
the optimal 
approximating density in a family ${\cal F}$ is 
$q^\star=\underset{q\in {\cal F}}{\mbox{argmax}}\;{\cal L}(q)$, 
where 
\[
{\cal L}(q)=\int \log\frac{g(\thetavec)}{q(\thetavec)}q(\thetavec)\mbox{d}\thetavec
=E_{q}\left[\log g(\thetavec)-\log q(\thetavec)\right]
\]
is called the 
evidence lower bound (ELBO).

In many applications, the family ${\cal F}$ is
specific to the statistical model being estimated. 
However, generic fixed form approximations that can 
be employed to estimate a wide range of statistical models are also used. 
These have densities $q_\lambda(\thetavec)$ with variational parameters $\lambdavec$ that 
fully specify the density, so that $q^\star(\thetavec)=q_{\lambda^\star}(\thetavec)$ with $\lambdavec^\star=\underset{\lambdavec}{\mbox{argmax}}\;{\cal L}(q_\lambda)$. 
When combined
with generic optimization algorithms, the
approach is called ``black box'' VI~\citep{ranganath14}. In this paper we refer to solving this optimization problem as ``calibration'' of the VA.

It is common to use stochastic gradient ascent (SGA) optimization, as
we do here. In SGA, given an initial, 
value $\lambdavec^{(0)}$, the variational parameters are updated
recursively as
\[
\lambdavec^{(s+1)}=\lambdavec^{(s)}+\deltavec^{(s)}\circ \left.\widehat{\nabla_\lambda {\cal L}(q_\lambda)}\right\vert_{\lambda=\lambda^{(s)}}\,,\;\; s=0,1,2,\ldots\,.
\]
Here, $\deltavec^{(s)}$ is a vector of adaptive step sizes, the operator
``$\circ$'' is the element-wise product, and $\widehat{\nabla_\lambda {\cal L}(q_\lambda)}$ is an unbiased estimator of the gradient evaluated at $\lambdavec=\lambdavec^{(s)}$. 
A low variance
gradient estimator is the main requirement of effective SGA,
and  the ``re-parameterization trick''
is a particularly effective means of obtaining one. Here, 
the model parameters are re-parameterized as $\thetavec=h(\varepsilonvec,\lambdavec)\sim q_\lambda$, where $h$ is a deterministic function and $\varepsilonvec$ is distributed with density $f_\varepsilon$ that is invariant to $\lambdavec$. Then the ELBO is ${\cal L}(q_\lambda)=E_{f_\varepsilon}\left[
\log g(h(\varepsilonvec,\lambdavec)) - \log q_\lambda (h(\varepsilonvec,\lambdavec))
\right]$ and its gradient is 
\begin{eqnarray}
\nabla_\lambda {\cal L}(q_\lambda) &=&
E_{f_\varepsilon}\left[\nabla_\lambda
\log g(h(\varepsilonvec,\lambdavec)) - \nabla_\lambda \log q_\lambda (h(\varepsilonvec,\lambdavec)) \right] \nonumber \\
&= &E_{f_\varepsilon}\left[
\left.\frac{\partial h(\varepsilonvec,\lambdavec)}{\partial \lambdavec}\right.^\top \left( \nabla_\theta
\log g(\thetavec) - \nabla_\theta \log q_\lambda (\thetavec)\right) \right] 
\,,\label{eq:reparmgrad}
\end{eqnarray}
see~\cite{ong+ns16}. (The
notational conventions for  
derivatives used in this paper are outlined in Part~A of
the Online Appendix).
Equation~\eqref{eq:reparmgrad} is widely called a ``re-parameterization gradient'' and
a low variance unbiased estimate of it
is obtained by generating draws
$\varepsilonvec\sim f_\varepsilon$, computing $\thetavec=h(\varepsilonvec,\lambdavec)$ and then evaluating the terms in the parentheses at these values. It is often the case that only a single
draw of $\varepsilonvec$ is sufficient to produce a low noise gradient estimator, as we
do here.
Thus, to use SGA with the re-parameterization trick requires selecting a VA with density $q_\lambda$ that admits a
re-parameterization $\{f_\varepsilon, h\}$ from which generation and computation of~\eqref{eq:reparmgrad} is fast.

\subsection{Copula model variational approximations}
Copula models are scalable and tractable representations of 
multivariate distributions
and have been explored previously as VAs
by~\cite{han+ldc16,tran2015,smith2020trans,gunawan2021} and~\cite{chi21}.
A copula model  VA has density 
\begin{equation}
q_\lambda(\thetavec)=c(\uvec;\pivec)\prod_{i=1}^m q_{\lambda_i}(\theta_i)\,,\label{eq:copmod}
\end{equation}
where $\uvec=(u_1,\ldots,u_m)^\top$, $q_{\lambda_i}(\theta_i)$ is the 
marginal density of $\theta_i$ that is selected arbitrarily, and $u_i=Q_{\lambda_i}(\theta_i)=
\int_{-\infty}^{\theta_i}q_{\lambda_i}(s)\mbox{d}s$.
The function $c(\uvec;\pivec)$ is a called the ``copula density'' and is a well-defined density function on the
unit cube $[0,1]^m$ with uniform marginals and parameters $\pivec$.
Typically, the copula is selected from a list, 
with 
vine~\citep{czado2019} or elliptical~\citep{fang2002} copulas 
the most popular in high dimensions. Thus, a copula model VA is usually specified
by choices for $q_{\lambda_1},\ldots,q_{\lambda_m}$ and $c(\cdot;\pivec)$, with
variational parameters
$\lambdavec=(\pivec^\top,\lambdavec_1^\top,\ldots,\lambdavec_m^\top)^\top$. 

However, two problems
can arise when using copula models as high-dimensional VAs.  
First, an efficient re-parameterization
$\{f_\epsilon,h\}$ may be difficult to identify.
Second, the derivative $\nabla_\theta \log q_\lambda(\thetavec)$
is required to evaluate
the re-parameterization gradient 
at~\eqref{eq:reparmgrad}. By the chain rule,
\[
\nabla_\theta \log q_\lambda(\thetavec)=
\frac{\partial \uvec}{\partial \thetavec}^\top\left( \frac{\partial}{\partial \uvec} \log c(\uvec;\pivec)\right)^\top  +  \left( \frac{\partial}{\partial \theta_1}
\log q_{\lambda_1}(\theta_1),\ldots,
\frac{\partial}{\partial \theta_m}
\log q_{\lambda_m}(\theta_m)\right)^\top\,,
\]
where $\nabla_\theta \log q_\lambda(\thetavec)$ is a column vector. 
Here, the term
$\frac{\partial}{\partial \uvec} \log c(\uvec;\pivec)$ 
may be prohibitively slow to 
compute; 
for example, \cite{chi21} note this problem
for a vine copula.
However, ``implicit copulas'' defined through transformation are one class of copulas that can
avoid both problems, as now discussed.

\subsection{Implicit copula model variational approximations}\label{sec:icvi}
Consider a continuous random vector $\psivec=(\psi_1,\ldots,\psi_m)^\top$ with known
parametric distribution $F(\psivec;\pivec)$, marginals $F_1(\psi_1;\pivec_1),\ldots,F_m(\psi_m;\pivec_m)$, and parameters $\pivec=\cup_{i=1}^m \pivec_i$. Then
the copula of this distribution is unique and has density
\begin{equation}
c(\uvec;\pivec)=\frac{p(\psivec;\pivec)}{\prod_{i=1}^m p_i(\psi_i;\pivec_i)}\,,\label{eq:icop}
\end{equation}
where $p(\psivec;\pivec)=\frac{\partial}{\partial \psivec}F(\psivec;\pivec)$, 
$p_i(\psi_i;\pivec_i)=\frac{d}{d\psi_i}F_i(\psi_i;\pivec_i)$, and $\psi_i=F_i^{-1}(u_i;\pivec_i)$.
The copula parameters $\pivec$ are constrained to ensure they
are identified in the copula density at~\eqref{eq:icop}, which is called
an implicit copula. (Note that in~\cite{smith2020trans} the constrained values
are denoted $\widetilde{\pivec}$ to distinguish them from the uncontrained
values $\pivec$, although we do not do so here.)
Different choices for $F(\psivec;\pivec)$
produce different copula families. 
These include Gaussian copulas, t copulas, elliptical copulas, 
skew t copulas, factor copulas and copula processes; \cite{smith2021} provides an overview of the broad class of implicit copulas.

An implicit copula can also be defined through the element-wise transformations
\[
\psi_i=F_i^{-1}\left(Q_{\lambda_i}(\theta_i);\pivec_i\right)\equiv k_i(\theta_i)\,,\; \mbox{ for }i=1,\ldots,m\,.
\]
These are strictly increasing functions because
$F_i$ and $Q_{\lambda_i}$ are also.
If $k_i^\prime(\theta_i)=\frac{d}{d\theta_i}k_i(\theta_i)$,
then by a change of variables the marginal density of the 
VA is simply 
\begin{equation}
	q_{\lambda_i}(\theta_i)=p_i(\psi_i;\pivec_i)k_i^\prime(\theta_i)\,.
	\label{eq:qmargin}
\end{equation}
Thus, from~\eqref{eq:copmod}--\eqref{eq:qmargin} the joint density of the implicit copula model VA is
\begin{eqnarray}
	q_\lambda(\thetavec)&=&c(\uvec;\pivec)\prod_{i=1}^m q_{\lambda_i}(\theta_i)=\frac{p(\psivec;\pivec)}{\prod_{i=1}^m p_i(\psi_i;\pivec_i)}\prod_{i=1}^mp_i(\psi_i;\pivec_i)k_i^\prime(\theta_i)\nonumber \\
	&= & p(\psivec;\pivec)\prod_{i=1}^m k_i^\prime(\theta_i)\,.
	\label{eq:qjoint}
\end{eqnarray}
Computing the re-parameterization gradient using the expression at~\eqref{eq:qjoint} can be 
much simpler than that at~\eqref{eq:copmod}
because it avoids evaluation of the term $\frac{\partial}{\partial \uvec} \log c(\uvec;\pivec)$. 
To see this, if $k_i^{\prime\prime}(\theta_i)=\frac{d^2}{d \theta_i^2} k_i(\theta)$, then
the derivative
\begin{equation}
\nabla_\theta \log q_\lambda(\thetavec)= \frac{\partial \psivec}{\partial \thetavec}^\top\left(\frac{\partial}{\partial \psivec}
\log p(\psivec;\pivec)\right)^\top + \left(
\frac{k_1^{\prime\prime}(\theta_1)}{k_1^{\prime}(\theta_1)},
\ldots,\frac{k_m^{\prime\prime}(\theta_m)}{k_m^{\prime}(\theta_m)}
\right)^\top\,,\label{eq:gradlogq}
\end{equation}
where $\frac{\partial \psivec}{\partial \thetavec}=\mbox{diag}\left(k_1^\prime(\theta_1),\ldots,
k_m^\prime(\theta_m)\right)$ is a diagonal matrix. This 
expression, and thus the re-parameterization gradient, 
is fast to compute whenever $k_i^\prime$, $k_i^{\prime \prime}$ and $\frac{\partial}{\partial \psivec}
\log p(\psivec;\pivec)$ are also. 

A copula model is usually defined from~\eqref{eq:copmod} by selecting 
the marginals $q_{\lambda_1},\ldots,q_{\lambda_m}$ and copula $c$. 
However, the transformation representation motivates another way
to define an implicit copula model by selecting 
$k_1,\ldots,k_m$ and $p(\psivec;\pivec)$ instead. These 
can be chosen so that~\eqref{eq:gradlogq} and the re-parameterization 
gradient at~\eqref{eq:reparmgrad} are fast to compute in high dimensions. 
In particular, picking each transformation $k_i$ directly
avoids the need to evaluate either $F_i$ or $Q_{\lambda_i}$, which can 
be slow or difficult.
In principle, any tractable distribution can be adopted for $\psivec$, with~\cite{smith2021} 
outlining
a wide range of choices and their respective implicit copulas.
In the next section we detail how to implement variational
inference with a sub-class of elliptical copulas, but discuss 
the potential of other choices in Section~\ref{sec:discuss}.


%
%
%

\section{Learnable elliptical copula model VA}\label{sec:lVA}
\subsection{Definition}
For each $k_i$ we select a parametric monotonically increasing function denoted as $k_{\widetilde{\gamma}_i}$ with parameter vector $\widetilde{\gammavec}_i$, so that  $\lambdavec=\{\pivec,\widetilde{\gammavec}_1,\ldots,
\widetilde{\gammavec}_m\}$. \cite{smith2020trans} suggest using
the Yeo-Johnson (YJ)
and the inverse G-and-H (iGH) transformations for $k_{\widetilde{\gamma}_i}$, because they are popular
in data analysis as transformations 
from asymmetric to symmetric distributions.
(The G-and-H
transformation is from symmetry, which is why its inverse is used.) 
However, both transformations
produce marginals $q_{\lambda_i}(\theta_i)$ that 
are not invariant to changes in location and scale,
as demonstrated in Section~\ref{sec:flex}.
We therefore introduce location and scale parameters $\mu_i$ and $\sigma_i$ 
directly for $\theta_i$, so that 
\begin{equation}
	\psi_i=k_{\widetilde{\gamma}_i}(\theta_i)= t_{\gamma_i}\left(\frac{\theta_i-\mu_i}{\sigma_i}\right)\,,\label{eq:ktransf}
\end{equation}
where $t_{\gamma_i}$ is either the YJ or iGH transformation, or compositions 
of the two.
The transformation $k_{\widetilde{\gamma}_i}$
has parameters $\widetilde{\gammavec}_i=(\gammavec_i^\top,\mu_i,\sigma_i)^\top$
and is fast to implement within SGA methods.

An elliptical distribution is employed for $\psivec$, in which 
case~\eqref{eq:icop} is the density of an elliptical copula studied by~\cite{fang2002}.
We consider
the sub-class of 
elliptical distributions that can be written as a scale mixture of normals with
generative representation
\begin{equation}
	\psivec=\sqrt{W}\bm{X}\,, \label{eq:stochasticrep}
\end{equation}
where
$\bm{X}\sim N_m(\bm{0},\Sigma)$ and $W\sim F_W(\cdot;\omegavec)$ are 
independent. In Section~\ref{sec:reparamt} it is discussed
how this representation  can be used to  implement efficiently
the SGA algorithm with the reparameterization trick.
The density of an elliptical distribution with location zero and scale matrix $\Sigma$ is  
\[
p(\psivec;\pivec)=|\Sigma|^{-1/2}\widetilde{g}_{m,\omega}\left(\psivec^\top \Sigma^{-1}\psivec\right)\,,
\]
where $\widetilde{g}_{m,\omega}(x)={\cal K}_{m,\omega}g_\omega(x)$, the function $g_\omega:[0,\infty)\rightarrow (0,\infty)$ with parameters
$\omegavec$, $\pivec=(\omegavec^\top,\mbox{vech}(\Sigma)^\top)^\top$, and
${\cal K}_{m,\omega}=\Gamma(\frac{m}{2})/(\pi^{\frac{m}{2}}\int_{0}^\infty t^{\frac{m}{2}-1}g_\omega(t)dt)$
is a constant.
We set the location to zero and $\Sigma$ to be a correlation matrix 
to identify their values, which is necessary because
$\mu_i,\sigma_i$ are introduced in the transformation at~\eqref{eq:ktransf}.

While all elliptical distributions
are closed under marginalization (i.e. the marginal of an elliptical distribution is
also an elliptical distribution), only a sub-class have
marginals of the same parametric family, which is a property called ``consistency'' by~\cite{kano94}. This author shows that elliptical distributions with the
stochastic representation at~\eqref{eq:stochasticrep} are consistent whenever the
distribution of $W$ is not a function of $m$. In this case
the marginal has density  
$p_i(\psi_i;\pivec_i)=\widetilde{g}_{1,\omega}(\psi_i^2)$ with $\pivec_i=\omegavec$, which 
can be used to define the implicit copula density at~\eqref{eq:icop}. Inconsistent 
elliptical distributions can also be used in our framework, where $\psi_i=\sqrt{W}X_i$
has density of a different parametric form.
Table~\ref{tab:ellip} gives four elliptical distributions that are 
scale mixtures of normals. They are the consistent Gaussian, t and Laplace distributions, and the inconsistent exponential power distribution.
For each, the function $\widetilde{g}_{m,\omega}$,
its derivative, $\omegavec$ and the distribution of $W$ are given.
These are necessary
to implement variational inference using their implicit copula models as 
VAs.

\begin{sidewaystable}[p]
	\caption{Four Elliptical Distributions with Scale Mixture of Normals Generative Representations}\label{tab:ellip}
	\begingroup
	\renewcommand{\arraystretch}{1.5}
	\begin{center}	
		\begin{tabular}{lllll}\hline \hline
			Distribution &$\widetilde{g}_{m,\omega}(x)$ &$\widetilde{g}_{m,\omega}^\prime(x)$ &$\omegavec$ &$W$ Distribution or Density \\ \hline
			\begin{tabular}{l} Gaussian\\ $\psivec \sim N_m(\bm{0},\Sigma)$ \end{tabular} 
			& $(2\pi)^{-m/2} \exp\left(-\frac{x}{2}\right)$ &$-\frac{1}{2}\widetilde{g}_{m,\omega}(x)$ &$\emptyset$ &Point Mass at 1\\
			& & & &\\
			\begin{tabular}{l} Symmetric Laplace\\ $\psivec \sim SL_m(\bm{0},\Sigma)$ \end{tabular} 
			 &\begin{tabular}{l}$\frac{2}{(2\pi)^{m/2}}\left(\frac{x}{2}\right)^{\nu/2}K_\nu(\sqrt{2x})$\\ with $\nu=(2-m)/2$\end{tabular} &\begin{tabular}{l}$c_1\left[\left(\frac{x}{2}\right)^{\frac{\nu-1}{2}}K_\nu'(\sqrt{2x})+\right.$\\
				$\left.\frac{\nu}{4}\left(\frac{x}{2}\right)^\frac{v-2}{2}K_\nu(\sqrt{2x})\right]$\end{tabular} &$\emptyset$ &$W\sim \mbox{Exp}(1)$\\
			& & & &\\
			\begin{tabular}{l} Multivariate $t$\\ $\psivec \sim t_m(\bm{0},\Sigma,\nu)$ \end{tabular} 
			&$\frac{\Gamma\left(\frac{\nu+m}{2}\right)}{\Gamma\left(\frac{\nu}{2}\right)
				(\pi\nu)^{m/2}}\left(1+\frac{x}{\nu}\right)^{-\left(\nu+m\right)/2}$
			&$\frac{-(\nu+m)}{2\nu}\left(1+\frac{x}{\nu}\right)^{-1}\widetilde{g}_{m,\omega}(x)$
			&$\nu>0$ &$W \sim \nu/\chi^2(\nu)$\\
				& & & &\\
			\begin{tabular}{l} Exponential Power\\ $\psivec \sim EP_m(\bm{0},\Sigma,\beta)$ \end{tabular}
			&$\frac{m\Gamma\left(\frac{m}{2}\right)}{\Gamma\left(1+\frac{m}{2\beta}\right)\pi^{\frac{m}{2}}2^{1+\frac{m}{2\beta}}}\exp\left(-\frac{1}{2}x^\beta\right)$ 
			&$-\frac{\beta}{2} x^{\beta-1} \widetilde{g}_{m,\omega}(x)$
			 &$0<\beta\leq 1$ 
			 &\begin{tabular}{l}$f_W(w)=\frac{2^{1+\frac{m}{2}-\frac{m}{2\beta}} \Gamma\left(1+\frac{m}{2}\right)}{\Gamma\left(1+\frac{m}{2\beta}\right)}$\\
			 $\;\;\mbox{ }\times w^{m-3}f_{S,\beta}(w^{-2};2^{1-\frac{1}{\beta}})$\end{tabular}  \\ 
			\hline \hline
		\end{tabular}
	\end{center}
	The distributions have density $p(\psivec;\pivec)=|\Sigma|^{-1/2}\widetilde{g}_{m,\omegavec}(\psivec^\top \Sigma^{-1}\psivec)$, where $\mbox{dim}(\psivec)=m$ and $\pivec=(\mbox{vech}(\Sigma)^\top,\omegavec^\top)^\top$, and stochastic representation
	as a scale mixture of normals. 
	In the table,
	$K_\nu$ denotes the modified Bessel function of the third kind, $\Gamma$ the gamma function, $\mbox{Exp}(1)$ an exponential distribution with
	parameter 1, $\chi^2(\nu)$ a chi-squared distribution with parameter $\nu$, 
	$f_{S,\alpha}(\cdot;\sigma)$ is the density of the positive stable distribution, and 
	the constant $c_1 = \frac{2}{(2\pi)^{m/2}}$. For details on the Symmetric Laplace distribution see~\citet[Ch.5]{kotz2001} and on the Exponential Power distribution see~\cite{gomez2008}.
	\endgroup
\end{sidewaystable}

From~\eqref{eq:qjoint} our proposed learnable VA has joint and marginal densities
\begin{eqnarray}
	q_\lambda\left(\bm{\theta}\right) &=& |\Sigma|^{-1/2}\widetilde{g}_{m,\omega}\left(\psivec^\top \Sigma^{-1}\psivec\right)\prod_{i=1}^{m}t_{\gamma_i}'\left(\frac{\theta_i-\mu_i}{\sigma_i}\right)\frac{1}{\sigma_i}\,,\label{Eq:VA}\\
	q_{\lambda_i}(\theta_i) &=& p_i(\psi_i;\pivec_i)t_{\gamma_i}'\left(\frac{\theta_i-\mu_i}{\sigma_i}\right)\frac{1}{\sigma_i}\,,\label{eq:EVAmargin}
\end{eqnarray}
where $p_i(\psi_i;\pivec_i)=\widetilde{g}_{1,\omega}(\psi_i^2)$ if 
the distribution for $\psivec$ is consistent.
Notice that by introducing $\mu_i$ and $\sigma_i$, the marginal density
for $\theta_i$ is flexible
in terms of its location and scale. 
In contrast, \cite{smith2020trans} and~\cite{gunawan2021} instead include a location and scale
for $\psi_i$ (and not for $\theta_i)$ which restricts the
flexibility of the marginal approximation as illustrated in Section~\ref{sec:flex} below.

\subsection{Invariance of the marginal approximation}\label{sec:flex}
The marginal
of the Gaussian copula model VA in~\cite{smith2020trans} has density
\[
q_i^{\mbox{\tiny SLN2020}}(\theta_i)=
\frac{1}{\sigma_{\psi_i}}\phi\left(\frac{\psi_i-\mu_{\psi_i}}{\sigma_{\psi_i}}\right)
t_{\gamma_i}^\prime(\theta_i)\,,
\]
where $\phi(x)$ is a $N(0,1)$ density.
The accuracy of this approximation varies with the 
location and scale of $\theta_i$, which is a poor property for a VA. 
The same issue arises for other choices of $p(\psivec;\pivec)$, including
elliptical or mixture distributions.
In contrast, the marginal approximation at~\eqref{eq:EVAmargin}
is invariant to the location and scale of $\theta_i$ by definition.

To demonstrate the impact on accuracy, 
we compare the ability of the marginal of $q_i^{\mbox{\tiny SLN2020}}$ and
the density at~\eqref{eq:EVAmargin} with $p_i(\psi_i;\pivec_i)=\phi(\psi_i)$
to approximate
the skew normal (SN) distribution of~\cite{azzalini1996} for $\theta_i$. 
Both VAs are constructed using the YJ transformation for $t_{\gamma_i}$.
The SN parameters are set so that the Pearson skew coefficient is $0.8553$, but
the mean $\mu_{SN}$ and standard deviation $\sigma_{SN}$ vary.
Figure~\ref{fig:KLmargin} plots the two optimal VAs (i.e. those with the minimum KL
divergence) when (a)~$\mu_{SN}=15,\sigma_{SN}=1$, and
(b)~$\mu_{SN}=0,\sigma_{SN}=1$. Simply changing the location
$\mu_{SN}$ of the target distribution
increases the KL divergence from 0.013 to 0.105 for 
$q_i^{\mbox{\tiny SLN2020}}$, but does not affect the accuracy of
the approximation proposed here. 

For the same case,
Figure~\ref{fig:KLSN} plots the KL divergence of the two optimal VAs 
for a range of values of $\mu_{SN}$ and $\sigma_{SN}$. The accuracy of
$q_i^{\mbox{\tiny SLN2020}}$ varies substantially with $\mu_{SN}$ and $\sigma_{SN}$, and is much less accurate than that at~\eqref{eq:EVAmargin} 
for most values of $\mu_{SN},\sigma_{SN}$. 
As an upper bound, the KL divergence for the optimal Gaussian approximation is 
also plotted as a horizontal line. This is nested by both VAs and it can
be seen that $q_i^{\mbox{\tiny SLN2020}}$ is only slightly 
more accurate than a Gaussian VA for some SN target distributions.

\subsection{Factor correlation matrix}
If $\Sigma$ is unrestricted, then the number of elements increases 
quadratically with $m$, prohibiting usage of the VA in high dimensions.
A popular solution for Gaussian VAs is to adopt
a factor decomposition (also called a ``low rank plus diagonal'' decomposition in the machine 
learning literature) see~\cite{miller2017,ong+ns16,mishkin2018slang} and~\cite{zhou21} for recent examples. 
We follow this approach for the elliptical copula and set $\Sigma=BB^\top+D^2$,
where $B=\{b_{i,j}\}$ is an $(m\times K)$ factor loading matrix,
$D = \text{diag}(\bm{d})$ and $\bm{d}=\left(d_1,\dots,d_m\right)^\top$.
To aid identification we set the upper triangle of $B$ to zeros, and bound
$\dvec$ and the leading edge of $B$ to be positive. Alternative identifying restrictions may also be used as discussed in~\cite{zhou21} for Gaussian VAs.
However, unlike in~\cite{smith2020trans},
a complicating factor is that the positive definite matrix $\Sigma$ has a leading diagonal
of ones, so
the constraints $d_j^2+\sum_{k=1}^{K}b_{j,k}^2 = 1$, for $j=1,\ldots,m$ hold. 
A popular way of imposing such a constraint is to employ spherical co-ordinates; e.g. see~\cite{rebonato1999}. Here, this
is equivalent to restricting ${\bm{a}}_j = \left(a_{j,1},\dots,a_{j,K+1}\right)^\top = \left(d_j,b_{j,1},\dots,b_{j,K}\right)^\top$ to lie on the surface of the $(K+1)$-dimensional spherical space $\mathbb{S}^n = \{\bm{a}_j : \sum_{k=1}^{K+1}a_{j,k}^2 = 1\}$. We impose this spherical restriction by writing ${\bm{a}}_j$ in terms of the angles ${\bm{\varkappa}}_j = \left(\varkappa_{j,1},\dots,\varkappa_{j,K}\right)^\top$, such that
\begin{equation}\label{EQ:inverse_function}
	a_{j,k} = \begin{cases}
		\cos\varkappa_{j,1} & \text{for $k=1$},\\
		\cos\varkappa_{j,k}\prod_{i = 1}^{k-1}\sin\varkappa_{j,i} &   \text{for $1<k<K+1$},\\
		\prod_{i = 1}^{k-1}\sin\varkappa_{j,i} & \text{for $k=K+1$}\,,
	\end{cases}
\end{equation}
where $\varkappa_{j,k}\in(0,\pi)$ if $k<K$ and $\varkappa_{j,K}\in(0,2\pi)$. There is a one-to-one relationship between
the angles and $(B,\dvec)$.

One last complication is that our 
SGA optimization method is applicable to unconstrained
variational parameters. To address this, the angles are expressed using
a further monotonic transformation to the vectors
 ${{\bm{\tau}}}_j = \left({\tau}_{j,1},\dots,{\tau}_{j,K}\right)^\top\in \mathbb{R}^K$, with elements 
 $\tau_{j,k}=\Phi_1^{-1}(\varkappa_{j,k}/\pi)$ for $k<K$ and $\tau_{j,K}=\Phi_1^{-1}({\varkappa}_{j,K}/2\pi)$. Finally, the
 complete variational parameter vector is $\bm{\lambda} = \left(\bm{\mu}^\top,\bm{\sigma}^\top,\bm{\gamma}^\top,\bm{\tau}^\top,\omegavec^\top\right)^\top$, with $\bm{\mu} = \left(\mu_1,\dots,\mu_m\right)^\top$, $\bm{\tau} = \left(\bm{\tau}_1^\top,\dots,\bm{\tau}_m^\top\right)^\top$, $\bm{\sigma} = \left(\sigma_1,\dots,\sigma_m\right)^\top$, $\bm{\gamma} = \left(\bm\gamma_1^\top,\dots,\bm\gamma_m^\top\right)^\top$ and $\omegavec$ is dependent on 
 the choice of elliptical distribution (which is simply $\omegavec=\emptyset$ for the Gaussian
 and Laplace distributions). 
Using this parameterization, the derivative at~\eqref{eq:gradlogq} is 
fast to compute analytically; see Appendix~\ref{app:A}. 

\subsection{Application of the re-parametrization trick}\label{sec:reparamt}
For our transformation-based implicit copula
\begin{equation}
	\bm{\theta} = h\left(\bm{\varepsilon},\bm{\lambda}\right) = \left(\mu_1+\sigma_1t_{\gamma_1}^{-1}(\psi_1),\dots,\mu_m+\sigma_mt_{\gamma_m}^{-1}(\psi_m)\right)^\top\,.
\end{equation}
To employ the re-parameterization trick, $p(\psivec;\pivec)$ 
must have a generative
representation based on a random vector $\varepsilonvec$ that is 
unrelated to $\lambdavec$. The scale mixture
of normals at~\eqref{eq:stochasticrep} provides this representation, as detailed in
Algorithm~\ref{alg:smn}.

\begin{algorithm}
	\caption{\em (Generating $\thetavec$ from the elliptical copula model VA)}
	\label{alg:smn}
	\begin{itemize}
		\item[1.] Generate independently $\zvec\sim N_m(\bm{0},I)$, $\epsilonvec\sim N_m(\bm{0},I)$, $u\sim \mbox{Uniform}(0,1)$ and set
		$\varepsilonvec=\left(\bm{z}^\top,\bm{\epsilon}^\top,u\right)^\top$
		\item[2.] Set $w=F_{W}^{-1}(u;\omegavec)$ where $F_{W}^{-1}$ is the quantile function of
		$W$
		\item[3.] Set $\psivec=\sqrt{w}\left(B\zvec+D\epsilonvec\right)$
		\item[4.] Set $\bm{\theta} = h\left(\bm{\varepsilon},\bm{\lambda}\right)$
	\end{itemize}
\end{algorithm}
In this algorithm, 
Step~2 is unnecessary for the Gaussian distribution for $\psivec$, while for other
elliptical distributions $F_W^{-1}$ can
be evaluated either analytically or numerically. 
Because $W$ is a scalar,
the computational burden of this step does not increase with the dimension $m$ of the target 
distribution. 
All other steps are computationally inexpensive.
\section{Examples}\label{sec:examples}
We use two examples to illustrate the improvement that adopting the VA at~\eqref{Eq:VA}
can provide.

\subsection{Example~1: mixed logistic regression}
We re-estimate the mixed logistic regression example outlined in~\cite{smith2020trans}. This employs a longitudinal dataset 
on 500 subjects over 7 years with 8 fixed effects (including an intercept) and one subject-based $N(0,\exp(2\zeta))$ random effect. 
The exact posterior can be computed using MCMC methods, so that
the accuracy of any VA can be judged. The posterior of random effect
values are typically skewed, which is the case for this example.
There are 9 model parameters and 500 random effect values,
so that a VA to the augmented posterior
has $m=509$. 

We employ a t-copula model VA with density at~\eqref{Eq:VA} where
$\psivec\sim t_m(\bm{0},\Sigma,\nu)$, $\widetilde{g}_{m,\omega}$ is
specified in Table~\ref{tab:ellip}, and the invariant transformation at~\eqref{eq:ktransf} is used to form the implicit
copula. The Gaussian copula model VA outlined in~\cite{smith2020trans} (labeled SLN2020) is
used as a comparison. 
In both cases, we set $K=5$ factors and considered three transformations for $t_{\gamma_i}$: YJ, iGH
and a composition of two YJ transforms (Double-YJ). We also use
a t distribution VA, which corresponds to assuming identity 
transformations $t_{\gamma_i}(\theta_i)=\theta_i$ for $i=1,\ldots,m$ at~\eqref{eq:ktransf}.
When calibrating each of the three t-copula models and the t distribution 
using SGA, we found 
the variational parameter $\nu^*>29$, which indicates that a Gaussian 
copula dependence structure is optimal here. 
Figure~\ref{fig:polymoments} plots the first three posterior
moments of the VAs (vertical axes) against their true values (horizontal axes).
All approximations identify the posterior means and standard deviations well, but SLN2020 does a poor job of estimating the posterior skew in panel~(f) in comparison to our proposed VA in panel~(c).
To further visualize, Figure~\ref{fig:polyRE} plots the exact and variational marginal posteriors of 
three random effect values.
The increased accuracy of the VA at~\eqref{Eq:VA} compared to SLN2020 is clear. 

\subsection{Example~2: regularized correlation matrix}\label{sec:corrmat}

We consider estimation of a Gaussian copula model for multivariate data (not to be confused with the
copula model VA) with parameter correlation matrix $\Omega$. 
In high dimensions either $\Omega$ or $\Omega^{-1}$
is often assumed to be sparse or patterned; e.g. see~\cite{oh2017modeling}.
However, in some applications
this may be unreasonable, so that $\Omega$ is a full matrix and a regularized estimator
is used. Below we suggest regularization of $\Omega$ based on 
a spherical coordinate parameterization, and then apply our VI approach 
for estimation. We note that the problem of estimating
a regularized correlation matrix also arises in random effects~\citep{danaher2020}, 
multivariate probit~\citep{zhang2021large} and other statistical models.

\subsubsection{The Statistical Model}
Denote as $\bm{y}_i = \left(y_{i,1},\dots,y_{i,r}\right)^\top$ the $i$th observation of an $r-$dimensional vector of data, and $\bm{y} = \left(\bm{y}_1^\top,\dots,\bm{y}_N^\top\right)^\top$. The Gaussian copula model for this data has likelihood
\begin{equation}
	p(\bm{y}|\bm{\theta}) = \frac{1}{|\Omega|^{N/2}}\prod_{i = 1}^{N}\exp\left[-\frac{1}{2}\bm{x}_i^\top\left(\Omega^{-1}-I_r\right)\bm{x}_i\right]\prod_{j=1}^r g_j(y_{i,j})\,,\label{eq:copeg}
\end{equation}
where $\bm{x}_i = \left(\Phi^{-1}(u_{i,1}),\dots,\Phi^{-1}(u_{i,r})\right)^\top$, $u_{i,j} = G_j(y_{i,j})$ and  $G_j,\,g_j$ denote the marginal distribution and density functions 
of $y_{i,j}$. 
We follow~\cite{pinheiro1996,rebonato1999} and~\cite{yoshiba2018maximum}, and
set $\Omega = LL^\top$, where $L=\{l_{i,j}\}$ is a lower 
triangular Cholesky factor with elements written in terms of spherical 
co-ordinates $\vartheta_{i,j}$ as
\[
l_{i,j} =
\left\{\begin{array}{cl}
	1 &\text{if} \ \ \ \ i=j=1\\
	\sin\left(\vartheta_{i,i-1}\right)\prod_{s=1}^{i-2}\sin\left(\vartheta_{i,s}\right) &\text{if} \ \ \ \ i=j>1\\
	\cos\left(\vartheta_{i,j}\right)\prod_{s=1}^{j-1}\sin\left(\vartheta_{i,s}\right)&\text{if} \ \ \ \ i>j\,\mbox{ and } i>1\\
	0 &\text{otherwise},\\
\end{array} \right.
\]
where $\prod_{s=1}^{0}\sin\left(\vartheta_{i,s}\right)\equiv 1$ and $\vartheta_{i,j}\in[0,\pi)$.
Denote $\bm{\vartheta}_i = \left(\vartheta_{i,1},\dots,\vartheta_{i,i-1}\right)^\top$, then
the angles $\bm{\vartheta} = \left(\bm{\vartheta}_2^\top,\dots,\bm{\vartheta}_r^\top\right)^\top$ provide a unique parameterization of $\Omega$. If $(X_1,\ldots,X_m)^\top\sim N(\bm{0},\Omega)$, each angle is 
related to a partial correlation as
\[
\cos\vartheta_{i,j}=\mbox{Corr}(X_i,X_j|X_1,\ldots,X_{j-1})\equiv \rho_{i,j|1:(j-1)}\,.
\]
Each angle is transformed to the real line by setting $\eta_{i,j} = \Phi^{-1}(\frac{\vartheta_{i,j}}{\pi})$. 
Each partial correlation $\rho_{i,j|1:(j-1)}=0$ iff
$\eta_{i,j}=0$, so that shrinkage of $\eta_{i,j}$ towards zero is equivalent
to shrinking the partial correlation to zero, representing conditional 
independence between $X_i$ and $X_j$. 

The parameters are stacked into the vector $\bm{{\eta}} = \left(\bm{{\eta}}_2^\top,\dots,\bm{{\eta}}_r^\top\right)^\top\equiv (\eta_1,\eta_2,\ldots,\eta_{r(r-1)/2})^\top$, with $\bm{{\eta}}_i = \left({\eta}_{i,1},\dots,{\eta}_{i,i-1}\right)^\top$. The 
horseshoe prior~\citep{carvalho2010} is used to
provide adaptive shrinkage towards zero (i.e. regularization), where 
\[
\eta_s|\xi,\chi_{s}~\sim N(0,\xi\chi_{s}),\ \  \chi_{s}|\nu_{s}\sim \mathcal{G}^{-1}\left(\frac{1}{2},\frac{1}{\nu_{s}}\right),\ \  \xi|\kappa\sim \mathcal{G}^{-1}\left(\frac{1}{2},\frac{1}{\kappa}\right),\ \  \nu_{1},\dots,\nu_{r(r-1)/2},\kappa\sim \mathcal{G}^{-1}\left(\frac{1}{2},20\right)\,,
\]
where $\mathcal{G}^{-1}(a,b)$ is an inverse gamma distribution with 
parameters $a$ and $b$.
The posterior densities of parameters that are regularized
using the horseshoe prior are funnel-shaped and difficult
to approximate~\citep{betancourt2015,ghosh2019}.
To address this, \cite{ingraham2017} suggest adopting the ``non-centered'' re-parameterization
\[
\eta_s = \tau_s\sqrt{\xi\chi_s}\,,\mbox{ for }s=1,\ldots,r(r-1)/2\,,
\]
which simplifies the geometry of the posterior.
Here, $\bm{\chi}^\top,\bm{\nu}^\top,\xi,\kappa$ are all positive, so we
further transform them to the real line using logarithmic transformations.
With this, the model parameter vector $\bm{\theta} = \left(\bm{\tau}^\top,\log \bm{\chi}^\top,\log \xi,\log \bm{\nu}^\top,\log \kappa\right)^\top$.
The posterior $p(\thetavec|\yvec)\propto p(\yvec|\thetavec)p(\thetavec)=g(\thetavec)$ and its gradient $\nabla_\theta
\log g(\thetavec)$ are available in closed form and are given
in Part~C of the Online Appendix. 

\subsubsection{U.S. income inequality data}
The copula model is used to model the dependence between changes in 
income inequality in U.S. states from 
1916 to 2018. 
Inequality is measured using the annual Gini coefficient, where higher values indicate higher income inequality. Data for the study were 
sourced from Mark Frank's webpage at
\underline{www.shsu.edu/eco\_mwf/inequality.html} using an approach
detailed in~\cite{frank09}. Due to missing observations the states of Alaska and Hawaii were removed, while the District of Columbia (Washington D.C.) was included.
Let $\mbox{GINI}_{t,j}$ denote
the Gini coefficient in year $t$ for state $j$, then we set 
$y_{t,j}=\mbox{GINI}_{t,j}-\mbox{GINI}_{t-1,j}$ and estimate $G_j$ in each state using a kernel density estimate (KDE). These are  non-Gaussian, so that a copula model is appropriate; see Figure~A1 in the Online Appendix. 
The copula data $u_{t,j}$ are then computed,  
and posterior inference for $\Omega$ evaluated for the two cases below.

\subsubsection*{Low dimensional case: two U.S. states}
The first estimation is for the $r=2$ most populous states of California and Texas. Here, 
$(\theta_1,\ldots,\theta_5)^\top=(\tau_1,\log \chi_1,  \log \xi, \log \nu_1,\log \kappa)^\top$ and 
we consider this case because in a low-dimensional setting the
posterior can be computed using MCMC methods, allowing comparison
with our VA. Figure~\ref{fig:CATX_mcmc} plots the univariate marginals $p(\theta_i|\yvec)$ and
bivariate marginals $p(\theta_i,\theta_j|\yvec)$ evaluated using MCMC.
The posteriors are unimodal, skewed and 
dependent-- precisely the type of distribution that can be captured well using an elliptical
copula model VA.
Figure~\ref{fig:CATX_copVA} plots the equivalent marginals of the t-copula model VA, with $t_{\gamma_i}$
being Double-YJ transforms and $K=5$ factors (i.e. $B$ is $5 \times 5$). These are given by~\eqref{eq:EVAmargin} and
bivariate slices of~\eqref{Eq:VA}.
The VA captures key features of the
posterior distribution well, including the location, skew and dependence. There is some
mild under-estimation 
of the posterior variance, which is a common feature of fixed-form VAs that may be 
rectified using a post-estimation adjustment~\citep{yu2021}.

\subsubsection*{Higher dimensional case: 49 U.S. states}
The second case is estimation for all $r=49$ U.S. states, so that $\Omega$ has $49(48/2)=1176$ 
unique elements and, when including the horseshoe hyper-parameters, $\thetavec$ has $m=3530$ elements. 
Table~\ref{tab:VAs_gini} summarizes the results for 20 different t-copula model VAs with differing numbers of columns $K$ for matrix $B$ and 
transformations $t_{\gamma_i}$ employed. 
When the identity transformation $t_{\gamma_i}(\theta_i)=\theta_i$ is used,
the VA is simply a t-distribution with a factor covariance matrix. 
When $K=0$ the scale matrix $\Sigma$ is diagonal and the
VA is a mean field approximation over each element of $\thetavec$. 

The number of variational parameters $|\lambdavec|$
and the time taken to complete 1,000 steps of the SGA algorithm are reported. We found 
15,000 steps is sufficient to calibrate each VA; for example, calibration 
took two hours for 
the t-copula model with $K=15$ and the YJ transformation for $t_{\gamma_i}$
using a low end desktop.
Also reported is the median lower bound
over the last 500 steps, labeled as $\overline{\mbox{LB}}$, with
higher values indicating more accurate calibration. By this measure the copula model VAs are all
much more accurate than a t-distribution, with the most accurate being the 
t-copula model with $K=10$, $15$ or $20$ and the YJ transformation for $t_{\gamma_i}$. 

\begin{table}[!htbp]
	\begin{center}
	\caption{Summary of 20 VAs to the Posterior of the Gaussian Copula Model for U.S. Inequality}	\label{tab:VAs_gini}%
	\begin{tabular}{lrrrrr}
		\hline\hline
		               &                                                 \multicolumn{5}{c}{Number of Columns in $B$}                                                 \\ \cline{2-6}
		$t_{\gamma_i}$ & \multicolumn{1}{c}{$K=0$} & \multicolumn{1}{l}{$K=5$} & \multicolumn{1}{l}{$K=10$} & \multicolumn{1}{l}{$K=15$} & \multicolumn{1}{l}{$K=20$} \\ \hline
		               & \multicolumn{5}{l}{{\em Number of Variational Parameters $|\lambdavec|$}}                                                                    \\ \cline{2-6}
		Identity       &                     7,061 &                    24,711 &                     42,361 &                     60,011 &                     77,661 \\
		iGH            &                    14,121 &                    31,771 &                     49,421 &                     67,071 &                     84,721 \\
		YJ             &                    10,591 &                    28,241 &                     45,891 &                     63,541 &                     81,191 \\
		Double YJ      &                    14,121 &                    31,771 &                     49,421 &                     67,701 &                     84,721 \\
		               & \multicolumn{5}{l}{{\em Time (Mins. per 1000 Steps)}}                                                                                        \\ \cline{2-6}
		Identity       &                       5.6 &                       5.9 &                        6.4 &                        8.3 &                       13.1 \\
		iGH            &                       5.6 &                       6.1 &                        6.6 &                        8.8 &                       13.7 \\
		YJ             &                       5.6 &                       5.9 &                        6.4 &                        8.3 &                       13.2 \\
		Double YJ      &                       5.5 &                       5.9 &                        6.5 &                        8.4 &                       13.2 \\
		               & \multicolumn{5}{l}{$\overline{\mbox{LB}}$}                                                                                                   \\ \cline{2-6}
		Identity       &                     416.6 &                     466.3 &                      534.8 &                      521.2 &                      530.9 \\
		iGH            &                     596.7 &                     480.4 &                      657.5 &                      565.0 &                      610.4 \\
		YJ             &                     650.0 &                     657.1 &                      664.8 &                      663.8 &                      668.9 \\
		Double YJ      &                     637.5 &                     640.5 &                      648.8 &                      649.3 &                      655.3 \\ \hline\hline
	\end{tabular}%
\end{center}
 The computation was implemented in serial using MATLAB on a low end DELL desktop with an Intel i7-10700 CPU @ 2.9Ghz.
\end{table}%

Pairwise dependence in the fitted Gaussian copula can be measured by the matrix of Spearman correlations
$\Omega^s=\frac{ 6}{\pi}\mbox{arcsin}(\frac{1}{2}\Omega)$, where arcsin is applied element-wise
to $\frac{1}{2}\Omega$. Figure~A2 (see Online Appendix) plots the point estimate of $\Omega^s$ from the optimal t-copula model VA to the posterior. 
To further visualize the results, we select the five contiguous 
western states of AZ, CA, NV, OR and WA, and then 
plot the variational posteriors of the pairwise Spearmean correlations for these states in Figure~\ref{fig:west_spear}. This is undertaken 
for both the mean field t-distribution and the t-copula model VA with $K=15$ and the YJ transformation. The impact of employing the more flexible VA can be seen; for example, the variational posterior
for the correlation between WA and AZ is located at a lower value. 
 
\section{Discussion}\label{sec:discuss}
This paper aims to refine and extend the
copula VI approach outlined in~\cite{smith2020trans}, making five contributions.
First, 
Section~\ref{sec:cvi} clarifies 
why implicit copula models are an attractive choice of VA in high dimensions, 
compared to other copula models such as vine copula models. 
Second, an
adjustment to the transformations 
that define the implicit copula is proposed that increases the accuracy of the VA. 
Third, it is shown how to 
implement the re-parameterization trick for a sub-class of elliptical copulas. 
This is useful in practice because employing the
re-parameterization gradient at~\eqref{eq:reparmgrad} increases the efficiency 
of SGA greatly. Fourth, it is shown how spherical co-ordinates can be used 
to represent the factor decomposition of $\Sigma$, and constrain it to a correlation matrix efficiently.
Last, it is demonstrated how the proposed copula VI method can 
be used to approximate the complex posterior of a regularized correlation matrix.
This is a difficult posterior to evaluate exactly in high dimensions using standard MCMC methods.

Elliptical copula VI is an attractive black box method for
posteriors that are unimodal, have positive and negative dependencies between
parameters, and exhibit high levels of skew, as in Figures~\ref{fig:CATX_mcmc} and~\ref{fig:CATX_copVA}. 
\cite{gunawan2021} also employ implicit copula VI, but with a mixture of normals for 
$\psivec\sim F$. This further extends the dependence structure of the VA, 
although calibration is more computationally demanding because the re-parameterization trick cannot be used directly. Compositions of parametric transformations can be considered at~\eqref{eq:ktransf}, 
although experiments (unreported) suggest little improvement in the accuracy of the VA. 
However, a promising direction for future work is to consider multivariate transformations
of the implicit copula model to further enrich the VA, such as the sparse Householder transformations discussed by~\cite{hirt2019} and references
therein; although, the resulting VA will no longer be a copula model 
with a known dependence structure.
Another direction
is to consider a Gaussian or other process for $\psivec\sim F$. 
The resulting implicit copula is a ``copula process'' \citep{Wilson2010} which can admit a more
complex dependence structure for the VA, while still exploiting the computational advantages outlined
here. Finally, while we did not employ our method to estimate a Bayesian neural network, it can clearly 
be used to do so. For example, it
would be interesting to consider the Bayesian neural network in~\cite{hirt2019}, which 
uses the same horseshoe prior based regularization as is employed in our correlation matrix example. 
\appendix
\section{Gradients for implementation of SGA}\label{app:A}
This appendix provides the gradients to implement SGA with the re-parameterization trick
for the proposed elliptical copula model VA  at~\eqref{Eq:VA}. They do not vary by choice of target density, and MATLAB
code for their fast evaluation is provided in the Supplementary Material. 
 
\subsection*{Expression for $\nabla_\theta\log q_\lambda(\bm{\theta})$}
For the elliptical copula model VA, the gradient at~\eqref{eq:gradlogq} is
\begin{align*}
	\nabla_\theta\log q_\lambda(\bm{\theta}) &= 
	\frac{g_{m,\omega}^\prime (\psivec^\top \Sigma^{-1} \psivec)}{g_{m,\omega}(\psivec^\top \Sigma^{-1} \psivec)}2\frac{\partial \bm{\psi}}{\partial\bm\theta}\Sigma^{-1}\psivec
	+\left(\frac{t_{\gamma_1}''\left(\frac{\theta_1-\mu_1}{\sigma_1}\right)}{t_{\gamma_1}'\left(\frac{\theta_1-\mu_1}{\sigma_1}\right)\sigma_1},\dots,\frac{t_{\gamma_m}''\left(\frac{\theta_m-\mu_m}{\sigma_m}\right)}{t_{\gamma_m}'\left(\frac{\theta_m-\mu_m}{\sigma_m}\right)\sigma_m}\right)^\top,
\end{align*}
with 
$$\frac{\partial \bm{\psi}}{\partial\bm\theta} = \text{diag}\left(t_{\gamma_1}'\left(\frac{\theta_1-\mu_1}{\sigma_1}\right)/\sigma_1,\dots,t_{\gamma_m}'\left(\frac{\theta_m-\mu_m}{\sigma_m}\right)/\sigma_m\right).$$
Expressions for $t_{\gamma_j}^\prime$ and $t_{\gamma_j}^{\prime \prime}$ 
are provided in Table~1 of \cite{smith2020trans} for both the YJ and iGH transformations,
which are very fast to compute. The derivatives of compositions of these two transformations
are also easily computed using a trivial application of the chain rule. Expressions for $g_{m,\omega}$ and $g_{m,\omega}^\prime$ are given in Table~\ref{tab:ellip} of this paper for the four elliptical copulas. 
For the Gaussian copula the first ratio simplifies to
 $\frac{g_{m,\omega}^\prime (\psivec^\top \Sigma^{-1} \psivec)}{g_{m,\omega}(\psivec^\top \Sigma^{-1} \psivec)} = -1/2$, while for the t copula it simplifies to $\frac{g_{m,\omega}^\prime (\psivec^\top \Sigma^{-1} \psivec)}{g_{m,\omega}(\psivec^\top \Sigma^{-1} \psivec)} = -\frac{v+m}{2\nu}\left[1+\left(\frac{\psivec^\top \Sigma^{-1} \psivec}{\nu}\right)\right]^{-1}$.

\subsection*{Expression for $\frac{\partial \bm{\theta}}{\partial \bm{\lambda}}=\frac{\partial h(\bm{\varepsilon},\bm{\lambda})}{\partial \bm{\lambda}}$}
Denote $A = [\bm{d} \ \ B]$ to be the matrix of elements $a_{j,k}$,  $\bm{\varkappa} = \left(\bm{\varkappa}_1^\top,\dots,\bm{\varkappa}_m^\top\right)^\top$,  $P_1 = [\bm{0}_{K\times 1} \ \ I_{K}]^\top$ and $P_2 = \left(1,\bm{0}_{1\times K}\right)^\top$.
The derivatives of $\bm{\theta}=h(\bm{\varepsilon},\bm{\lambda})$ with respect to $\bm{\lambda} = \left(\bm{\mu}^\top,\bm{\sigma}^\top,\bm{\gamma}^\top,\bm{\tau}^\top,\omegavec^\top\right)^\top $ are 
\begin{align*}
\frac{\partial \bm{\theta}}{\partial \bm{\mu}} &= I_m\,,\hspace{1cm}
\frac{\partial \bm{\theta}}{\partial \bm{\sigma}} = \text{diag}\left(t_{\gamma_1}^{-1}(\psi_1),\dots,t_{\gamma_m}^{-1}(\psi_m)\right)\,,\\
\frac{\partial \bm{\theta}}{\partial \bm{\gamma}}& = \text{diag}\left(\sigma_1\frac{\partial t_{\gamma_1}^{-1}(\psi_1)}{\partial\bm{\gamma}_1},\dots,\sigma_m\frac{\partial t_{\gamma_m}^{-1}(\psi_m)}{\partial\bm{\gamma}_m}\right)\,,\\
 \frac{\partial \bm{\theta}}{\partial \bm{\tau}} &= \frac{\partial \bm{\theta}}{\partial \bm{\psi}}\frac{\partial \bm{\psi}}{\partial\bm{\varkappa}}\frac{\partial \bm{\varkappa}}{\partial\bm{\tau}}\,,\hspace{1cm}
\frac{\partial \bm{\theta}}{\partial \omegavec} =\frac{\partial \bm{\theta}}{\partial \bm{\psi}}\frac{\partial \bm{\psi}}{\partial\omegavec}\,, 
\end{align*}
where 
$$\frac{\partial \bm{\theta}}{\partial \bm{\psi}} = \text{diag}\left(\sigma_1\frac{\partial t_{\gamma_1}^{-1}(\psi_1)}{\partial\psi_1},\dots,\sigma_m\frac{\partial t_{\gamma_m}^{-1}(\psi_m)}{\partial\psi_m}\right)\,,\; \mbox{ and }
\frac{\partial \bm{\varkappa}}{\partial\bm{\tau}} = \text{blockdiag}\left(\frac{\partial \bm{\varkappa}_1}{\partial\bm{\tau}_1},\dots,\frac{\partial \bm{\varkappa}_m}{\partial\bm{\tau}_m}\right)\,,$$
with $\frac{\partial\bm{\varkappa}_j}{\partial\bm{\tau}_j} = \text{diag}\left(\phi(\tau_{j,1})\pi,\dots,\phi(\tau_{j,K-1})\pi,\phi(\tau_{j,K})2\pi\right)$, and
$$\frac{\partial\bm\psi}{\partial\bm\varkappa} = \sqrt{w}\left[(\bm{z}^\top P_1^\top)\otimes I_m+ \text{diag}\left(\bm{\epsilon}\right)\left(P_2^\top\otimes I_m\right)\right]K_{K+1,m}\frac{\partial A^\top}{\partial \bm{\varkappa}},$$
with  $\frac{\partial A^\top}{\partial \bm{\varkappa}} = \text{blockdiag}\left(\frac{\partial \bm{a}_1}{\partial \bm{\varkappa}_1},\dots,\frac{\partial \bm{a}_m}{\partial \bm{\varkappa}_m}\right)$ and
\[
\left\{\frac{\partial \bm{a}_i}{\partial \bm{\varkappa}_i}\right\}_{j,l} =
\left\{\begin{array}{cl}
	\cos\left(\varkappa_{i,j}\right)\cos\left(\varkappa_{i,l}\right)\prod_{s\in\{1,\dots,j-1\}\backslash l}\sin\left(\varkappa_{i,s}\right) &\text{If} \ \ \ \ l<j\mbox{ and } j<K+1\\
	-\prod_{s=1}^{j}\sin\left(\varkappa_{i,s}\right) &\text{If} \ \ \ \ l=j \mbox{ and }  j<K+1\\
	\cos\left(\varkappa_{i,l}\right)\prod_{s\in\{1,\dots,j-1\}\backslash l}\sin\left(\varkappa_{i,s}\right) &\text{If} \ \ \ \ l<j\mbox{ and }  j=K+1\\
	0 &\text{otherwise}\\
\end{array} \right..
\]
Here, $K_{K+1,m}$ denotes the relevant commutation matrix.  For details on the derivation of these expressions see Part~B of the Online Appendix.
The terms $\frac{\partial t_{\gamma_j}^{-1}(\cdot)}{\partial\bm{\gamma}_j}$ and $\frac{\partial t_{\gamma_j}^{-1}(\cdot)}{\partial{\psi}_j}$ are provided in Table 1 of \cite{smith2020trans} for both the YJ and iGH transformations.

For the Gaussian and Laplace distributions it is unnecessary to evaluate
the derivative $\frac{\partial \bm{\psi}}{\partial\omegavec}$ because $\omegavec=\emptyset$.
However, in general $\psivec=\sqrt{w}(B\zvec+D\epsilonvec)$ with $w=F_W^{-1}(u;\omegavec)$, 
so that
\[
\frac{\partial \bm{\psi}}{\partial\omegavec} = \left(B\bm{z}+D\bm{\epsilon}\right)\times\left\{
 \frac{1}{2\left(F_W^{-1}(u;\omegavec)\right)^{1/2}} 
 \left(\frac{\partial}{\partial \omegavec} F_W^{-1}(u;\omegavec)\right)\right\}
\]
For the $t$ distribution $\omegavec=\nu$, $F^{-1}_W(u;\omegavec)=\nu/F_X^{-1}(1-u;\nu)$ with $F_X$ the distribution function of $X\sim \chi^2(\nu)$, and $\frac{\partial}{\partial \bm{\omega}}F_W^{-1}(u;\bm{\omega})$ is computed numerically.

\section*{Supplementary Materials}
\begin{description}
	\item[Online Appendix:] provides further details on notation used, the derivatives, examples and MATLAB code. 
	\item[Code and Data Zip File:] provides MATLAB code and data to replicate our results, and apply the method to other datasets and/or models.
\end{description}

\singlespacing
\newpage
\bibliography{references}

\begin{thebibliography}{}

\bibitem[\protect\astroncite{Azzalini and Dalla~Valle}{1996}]{azzalini1996}
Azzalini, A. and Dalla~Valle, A. (1996).
\newblock The multivariate skew-normal distribution.
\newblock {\em Biometrika}, 83(4):715--726.

\bibitem[\protect\astroncite{Betancourt and Girolami}{2015}]{betancourt2015}
Betancourt, M. and Girolami, M. (2015).
\newblock Hamiltonian {M}onte {C}arlo for hierarchical models.
\newblock {\em Current trends in {B}ayesian methodology with applications},
  79(30):2--4.

\bibitem[\protect\astroncite{Carvalho et~al.}{2010}]{carvalho2010}
Carvalho, C.~M., Polson, N.~G., and Scott, J.~G. (2010).
\newblock The horseshoe estimator for sparse signals.
\newblock {\em Biometrika}, 97(2):465--480.

\bibitem[\protect\astroncite{Chi et~al.}{2021}]{chi21}
Chi, J., Ouyang, J., Zhang, A., Wang, X., and Li, X. (2021).
\newblock Fast copula variational inference.
\newblock {\em Journal of Experimental \& Theoretical Artificial Intelligence},
  0(0):1--16.

\bibitem[\protect\astroncite{Czado}{2019}]{czado2019}
Czado, C. (2019).
\newblock Analyzing dependent data with vine copulas.
\newblock {\em Lecture Notes in Statistics, Springer}.

\bibitem[\protect\astroncite{Danaher et~al.}{2020}]{danaher2020}
Danaher, P.~J., Danaher, T.~S., Smith, M.~S., and Loaiza-Maya, R. (2020).
\newblock Advertising effectiveness for multiple retailer-brands in a
  multimedia and multichannel environment.
\newblock {\em Journal of Marketing Research}, 57(3):445--467.

\bibitem[\protect\astroncite{Fang et~al.}{2002}]{fang2002}
Fang, H.-B., Fang, K.-T., and Kotz, S. (2002).
\newblock The meta-elliptical distributions with given marginals.
\newblock {\em Journal of Multivariate Analysis}, 82(1):1--16.

\bibitem[\protect\astroncite{Frank}{2009}]{frank09}
Frank, M.~W. (2009).
\newblock Inequality and growth in the united states: Evidence from a new
  state-level panel of income inequality measures.
\newblock {\em Economic Inquiry}, 47(1):55--68.

\bibitem[\protect\astroncite{Ghosh et~al.}{2019}]{ghosh2019}
Ghosh, S., Yao, J., and Doshi-Velez, F. (2019).
\newblock Model selection in {B}ayesian neural networks via horseshoe priors.
\newblock {\em Journal of Machine Learning Research}, 20(182):1--46.

\bibitem[\protect\astroncite{G{\'o}mez-S{\'a}nchez-Manzano
  et~al.}{2008}]{gomez2008}
G{\'o}mez-S{\'a}nchez-Manzano, E., G{\'o}mez-Villegas, M., and Mar{\'\i}n, J.
  (2008).
\newblock Multivariate exponential power distributions as mixtures of normal
  distributions with bayesian applications.
\newblock {\em Communications in Statistics—Theory and Methods},
  37(6):972--985.

\bibitem[\protect\astroncite{Gunawan et~al.}{2021}]{gunawan2021}
Gunawan, D., Kohn, R., and Nott, D. (2021).
\newblock Flexible variational {B}ayes based on a copula of a mixture of
  normals.
\newblock {\em arXiv preprint arXiv:2106.14392}.

\bibitem[\protect\astroncite{Han et~al.}{2016}]{han+ldc16}
Han, S., Liao, X., Dunson, D.~B., and Carin, L.~C. (2016).
\newblock Variational {G}aussian copula inference.
\newblock In Gretton, A. and Robert, C.~C., editors, {\em Proceedings of the
  19th International Conference on Artificial Intelligence and Statistics},
  volume~51, pages 829--838, Cadiz, Spain. JMLR Workshop and Conference
  Proceedings.

\bibitem[\protect\astroncite{Hirt et~al.}{2019}]{hirt2019}
Hirt, M., Dellaportas, P., and Durmus, A. (2019).
\newblock Copula-like variational inference.
\newblock In Wallach, H., Larochelle, H., Beygelzimer, A., d\textquotesingle
  Alch\'{e}-Buc, F., Fox, E., and Garnett, R., editors, {\em Advances in Neural
  Information Processing Systems}, volume~32.

\bibitem[\protect\astroncite{Ingraham and Marks}{2017}]{ingraham2017}
Ingraham, J. and Marks, D. (2017).
\newblock Variational inference for sparse and undirected models.
\newblock In {\em International Conference on Machine Learning}, pages
  1607--1616. PMLR.

\bibitem[\protect\astroncite{Kano}{1994}]{kano94}
Kano, Y. (1994).
\newblock Consistency property of elliptic probability density functions.
\newblock {\em Journal of Multivariate Analysis}, (1):139--147.

\bibitem[\protect\astroncite{Kotz et~al.}{2001}]{kotz2001}
Kotz, S., Kozubowski, T.~J., and Podg{\'o}rski, K. (2001).
\newblock {\em The Laplace Distribution and Generalizations}.
\newblock Springer.

\bibitem[\protect\astroncite{Miller et~al.}{2017}]{miller2017}
Miller, A.~C., Foti, N.~J., and Adams, R.~P. (2017).
\newblock Variational boosting: Iteratively refining posterior approximations.
\newblock In {\em International Conference on Machine Learning}, pages
  2420--2429. PMLR.

\bibitem[\protect\astroncite{Mishkin et~al.}{2018}]{mishkin2018slang}
Mishkin, A., Kunstner, F., Nielsen, D., Schmidt, M., and Khan, M.~E. (2018).
\newblock Slang: Fast structured covariance approximations for {B}ayesian deep
  learning with natural gradient.
\newblock {\em arXiv preprint arXiv:1811.04504}.

\bibitem[\protect\astroncite{Nelsen}{2006}]{nelsen06}
Nelsen, R.~B. (2006).
\newblock {\em An Introduction to Copulas.}
\newblock Springer-Verlag, New York, Secaucus, NJ, USA.

\bibitem[\protect\astroncite{Oh and Patton}{2017}]{oh2017modeling}
Oh, D.~H. and Patton, A.~J. (2017).
\newblock Modeling dependence in high dimensions with factor copulas.
\newblock {\em Journal of Business \& Economic Statistics}, 35(1):139--154.

\bibitem[\protect\astroncite{Ong et~al.}{2018}]{ong+ns16}
Ong, V. M.-H., Nott, D.~J., and Smith, M.~S. (2018).
\newblock {Gaussian variational approximation with factor covariance
  structure}.
\newblock {\em Journal of Computational and Graphical Statistics},
  27(3):465--478.

\bibitem[\protect\astroncite{Pinheiro and Bates}{1996}]{pinheiro1996}
Pinheiro, J.~C. and Bates, D.~M. (1996).
\newblock Unconstrained parametrizations for variance-covariance matrices.
\newblock {\em Statistics and computing}, 6(3):289--296.

\bibitem[\protect\astroncite{Ranganath et~al.}{2014}]{ranganath14}
Ranganath, R., Gerrish, S., and Blei, D. (2014).
\newblock {Black Box Variational Inference}.
\newblock In Kaski, S. and Corander, J., editors, {\em Proceedings of the
  Seventeenth International Conference on Artificial Intelligence and
  Statistics}, volume~33 of {\em Proceedings of Machine Learning Research},
  pages 814--822, Reykjavik, Iceland. PMLR.

\bibitem[\protect\astroncite{Rebonato and J{\"a}ckel}{1999}]{rebonato1999}
Rebonato, R. and J{\"a}ckel, P. (1999).
\newblock The most general methodology to create a valid correlation matrix for
  risk management and option pricing purposes.
\newblock {\em Journal of Risk}, 2(2):17--27.

\bibitem[\protect\astroncite{Rothman et~al.}{2008}]{rothman08}
Rothman, A.~J., Bickel, P.~J., Levina, E., and Zhu, J. (2008).
\newblock {Sparse permutation invariant covariance estimation}.
\newblock {\em Electronic Journal of Statistics}, 2:494 -- 515.

\bibitem[\protect\astroncite{Smith}{2021}]{smith2021}
Smith, M.~S. (2021).
\newblock Implicit copulas: An overview.
\newblock {\em Econometrics and Statistics}, Forthcoming.

\bibitem[\protect\astroncite{Smith et~al.}{2020}]{smith2020trans}
Smith, M.~S., Loaiza-Maya, R., and Nott, D.~J. (2020).
\newblock High-dimensional copula variational approximation through
  transformation.
\newblock {\em Journal of Computational and Graphical Statistics},
  29(4):729--743.

\bibitem[\protect\astroncite{Tran et~al.}{2015}]{tran2015}
Tran, D., Blei, D., and Airoldi, E.~M. (2015).
\newblock Copula variational inference.
\newblock In {\em Advances in Neural Information Processing Systems}, pages
  3564--3572.

\bibitem[\protect\astroncite{Wilson and Ghahramani}{2010}]{Wilson2010}
Wilson, A.~G. and Ghahramani, Z. (2010).
\newblock Copula processes.
\newblock In {\em Advances in Neural Information Processing Systems}, pages
  2460--2468.

\bibitem[\protect\astroncite{Yoshiba}{2018}]{yoshiba2018maximum}
Yoshiba, T. (2018).
\newblock Maximum likelihood estimation of skew-t copulas with its applications
  to stock returns.
\newblock {\em Journal of Statistical Computation and Simulation},
  88(13):2489--2506.

\bibitem[\protect\astroncite{Yu et~al.}{2021}]{yu2021}
Yu, X., Nott, D.~J., Tran, M.-N., and Klein, N. (2021).
\newblock Assessment and adjustment of approximate inference algorithms using
  the law of total variance.
\newblock {\em Journal of Computational and Graphical Statistics}, forthcoming.

\bibitem[\protect\astroncite{Zhang et~al.}{2021}]{zhang2021large}
Zhang, Z., Nishimura, A., Bastide, P., Ji, X., Payne, R.~P., Goulder, P.,
  Lemey, P., and Suchard, M.~A. (2021).
\newblock Large-scale inference of correlation among mixed-type biological
  traits with phylogenetic multivariate probit models.
\newblock {\em The Annals of Applied Statistics}, 15(1):230--251.

\bibitem[\protect\astroncite{Zhou et~al.}{2021}]{zhou21}
Zhou, B., Gao, J., Tran, M.-N., and Gerlach, R. (2021).
\newblock Manifold optimization-assisted gaussian variational approximation.
\newblock {\em Journal of Computational and Graphical Statistics}, Forthcoming.

\end{thebibliography}
\vspace{3cm}

\begin{figure}[!thb]
	\caption{Marginals of Different Gaussian Copula Model VAs}
	\begin{center}
		\includegraphics[trim = 2cm 0cm 2cm 0cm, clip,width=0.99\textwidth]{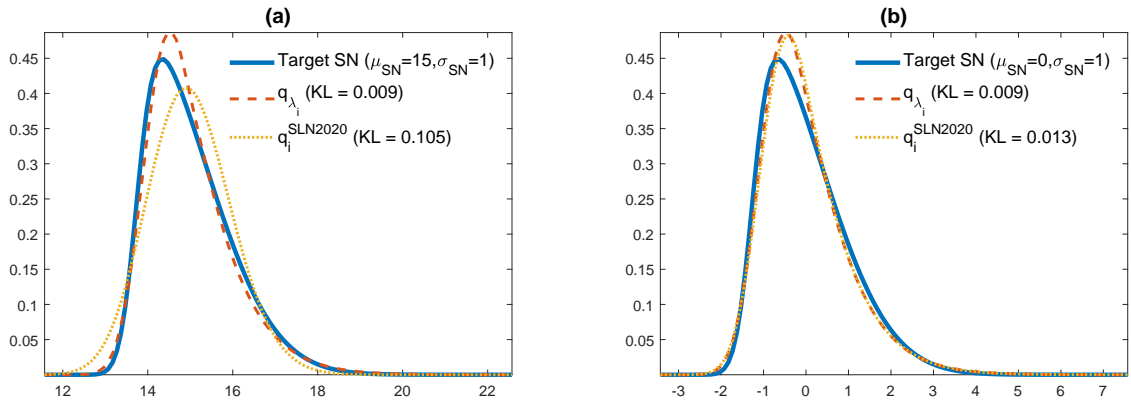}
	\end{center}
	Marginal densities $q_i^{\mbox{\tiny SLN2020}}$ (yellow dotted line) and $q_{\lambda_i}$ at~\eqref{eq:EVAmargin} (red dashed line) of optimal VAs
	to two skew normal (SN) target densities (blue solid line). Both target distributions have 
	the same Pearson skew of 0.8553 and standard deviation $\sigma_{SN}=1$. 
	Panel~(a) plots VA densities when the target mean is $\mu_{SN}=15$, and
	panel~(b) plots VA densities when the target mean is $\mu_{SN}=0$. Also 
	reported are the KL divergences for the optimal approximations in parentheses.
	\label{fig:KLmargin}
\end{figure}

\begin{figure}[!tb]
	\caption{Kullback-Leibler Divergence of Gaussian Copula Model VA Marginals}
	\begin{center}
		\includegraphics*[scale = 0.8]{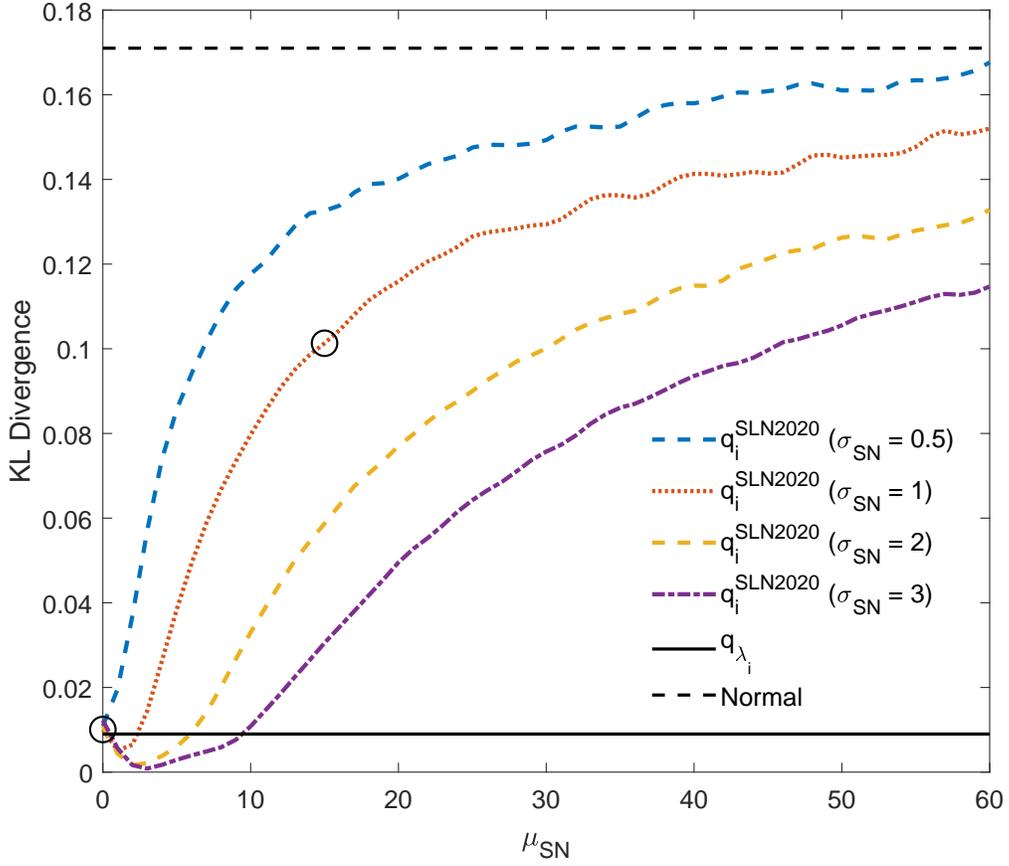}
	\end{center}
	KL divergence of optimal marginal approximations to SN target densities. 
	The SN distributions have a Pearson skew of 0.8553, but different means $\mu_{SN}$ and standard
	deviations $\sigma_{SN}$. Gaussian approximations have KL divergence
	of 0.172 for all target densities (black dashed line), and the approximation $q_{\lambda_i}$ at~\eqref{eq:EVAmargin} has KL divergence of 0.009 for all target densities (solid black line). The KL divergence of $q_i^{\mbox{\tiny SLN2020}}$ is plotted as curves for four values of $\sigma_{SN}$ and $0<\mu_{SN}<60$. 
	The two cases corresponding to Figure~\ref{fig:KLmargin}(a,b) are indicated with circles.
	\label{fig:KLSN}
\end{figure}

\begin{figure}[!tb]
	\caption{Accuracy of Posterior Moments for Mixed Logistic Regression Example}
	\begin{center}
		\includegraphics*[trim = 3cm 1cm 3cm 1cm,clip,width=\textwidth]{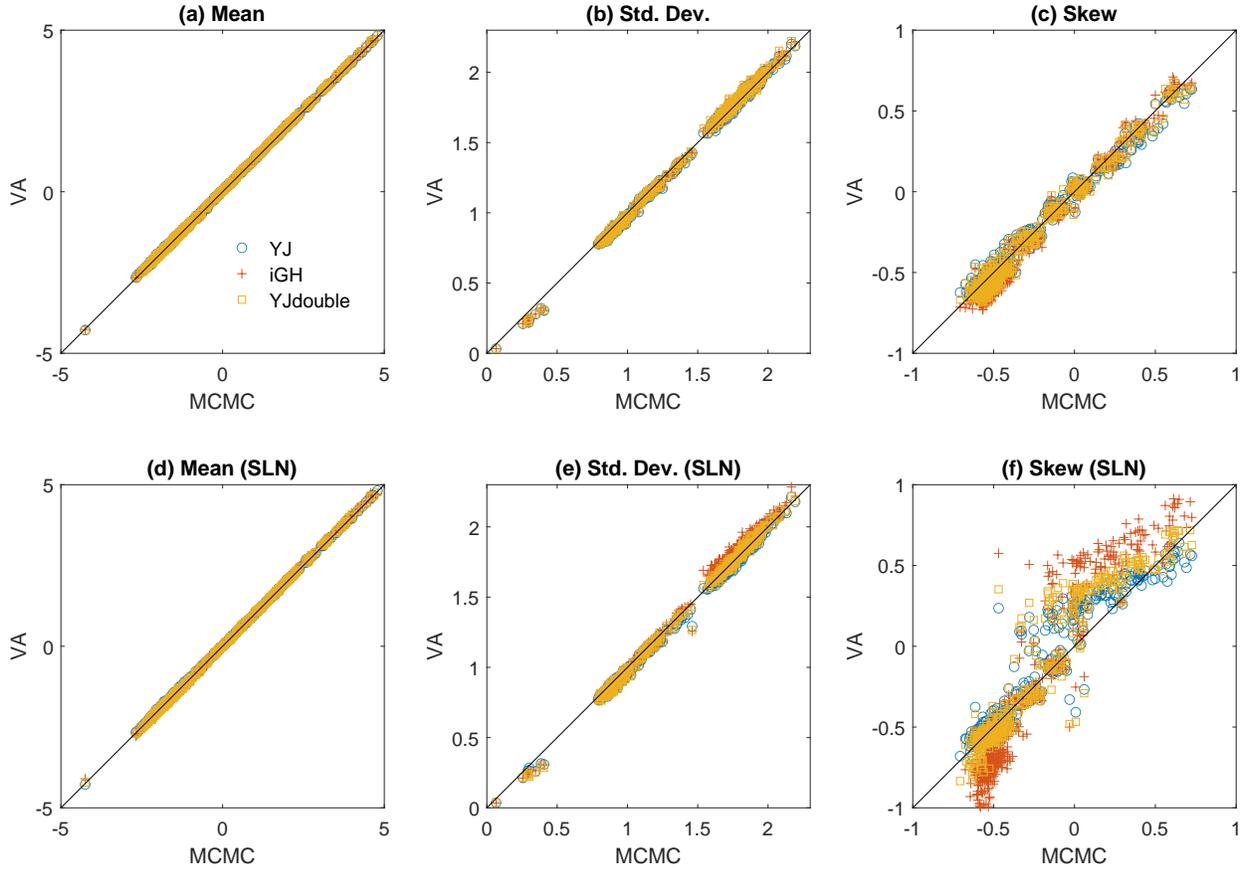}
	\end{center}
Panels (a)--(c) are for the t-copula model VA proposed here, and panels~(d)--(f) are for the Gaussian copula model VA in~\cite{smith2020trans}.
In each panel the exact posterior moment (computed using MCMC) is plotted on the horizontal axis, 
and moment of the VA on the vertical axis. The posterior means, standard deviations and Pearson's skew are plotted in the first, second and third columns, respectively. 
In each panel there is a scatter of $m=509$ 
points for each of three variational estimators that correspond to the YJ (blue circles),
iGH (red cross) and YJ-Double (yellow square) transformations for $t_{\gamma_i}$.
For panels~(a)--(e), the results
are very similar for the three transformations, so that most points are close to each other and hard to distinguish.
	\label{fig:polymoments}
\end{figure}

\begin{figure}[!tb]
	\caption{Posteriors for Three Random Effects in the Mixed Logistic Regression Example}
	\begin{center}
		\includegraphics*[width=\textwidth]{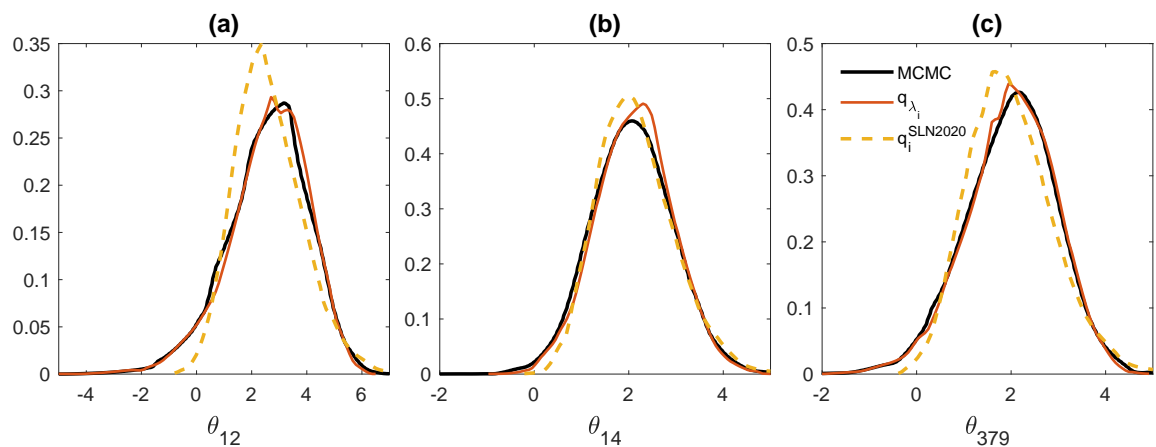}
	\end{center}
Marginal posterior densities for three random effects.  Each panel plots the exact posterior
computed by MCMC (black solid thick line), the t-copula model VA $q_{\lambda_i}$ at~\eqref{eq:qmargin} (red solid thin line) and that in~\cite{smith2020trans} $q_i^{\mbox{\tiny SLN2020}}$ (yellow dashed line).
Both approximations use the iGH transformation. 
	\label{fig:polyRE}
\end{figure}

\begin{figure}[!p]
	\caption{Exact Posterior for the Two U.S. States Inequality Example}
	\begin{center}
		\includegraphics*[width=\textwidth]{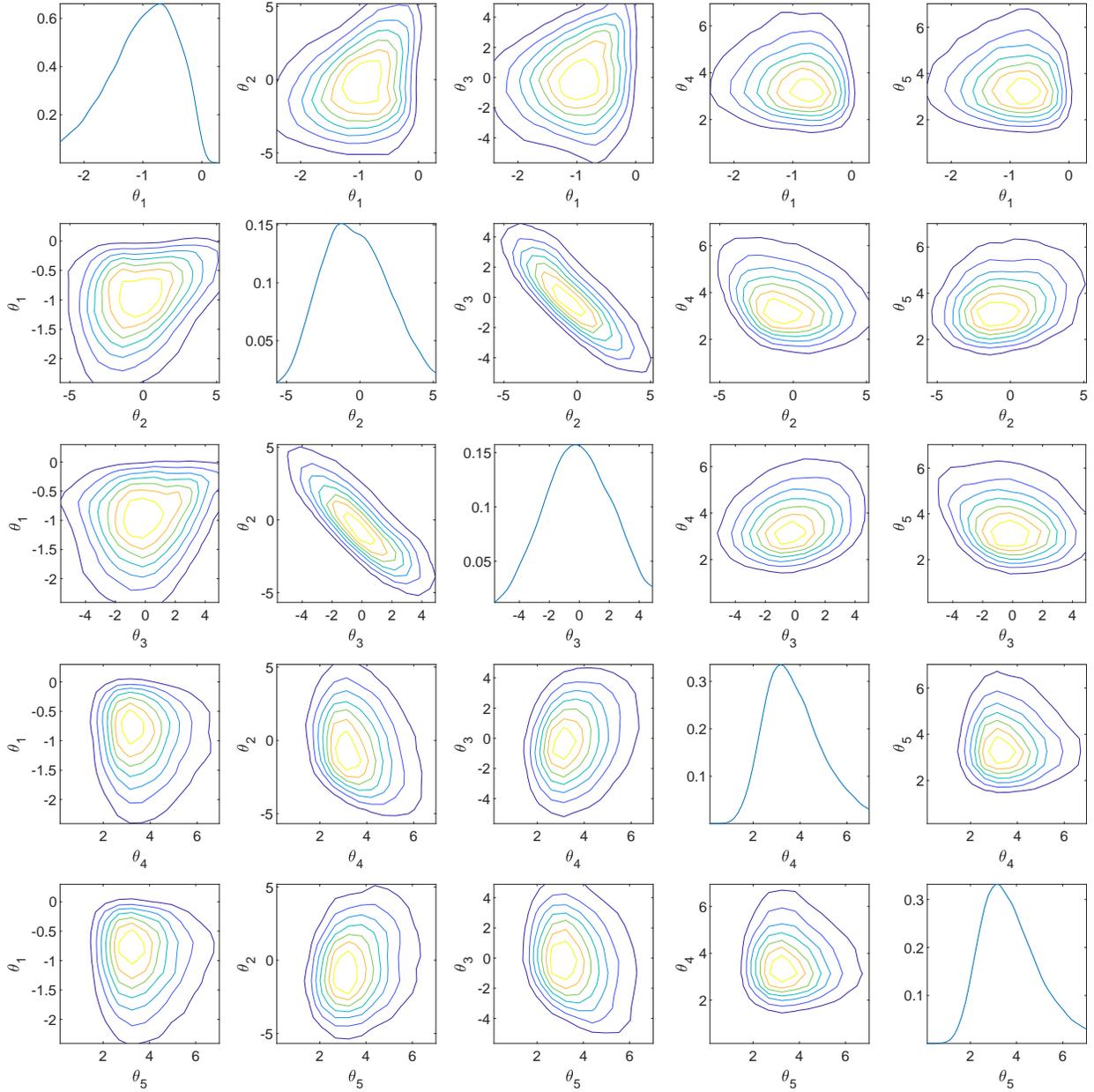}
	\end{center}
	Exact posterior of $(\theta_1,\ldots,\theta_5)^\top=(\tau_1,\log \chi_1, \log \nu_1, \log \xi, \log \kappa)^\top$ computed using MCMC. The univariate marginals $p(\theta_i|\yvec)$ are on the leading diagonals, and 
	the bivariate marginals $p(\theta_i,\theta_j|\yvec)$ are on the off-diagonals. 
	\label{fig:CATX_mcmc}
\end{figure}

\begin{figure}[!p]
	\caption{Variational Posterior for the Two U.S. States Inequality Example}
	\begin{center}
		\includegraphics*[width=\textwidth]{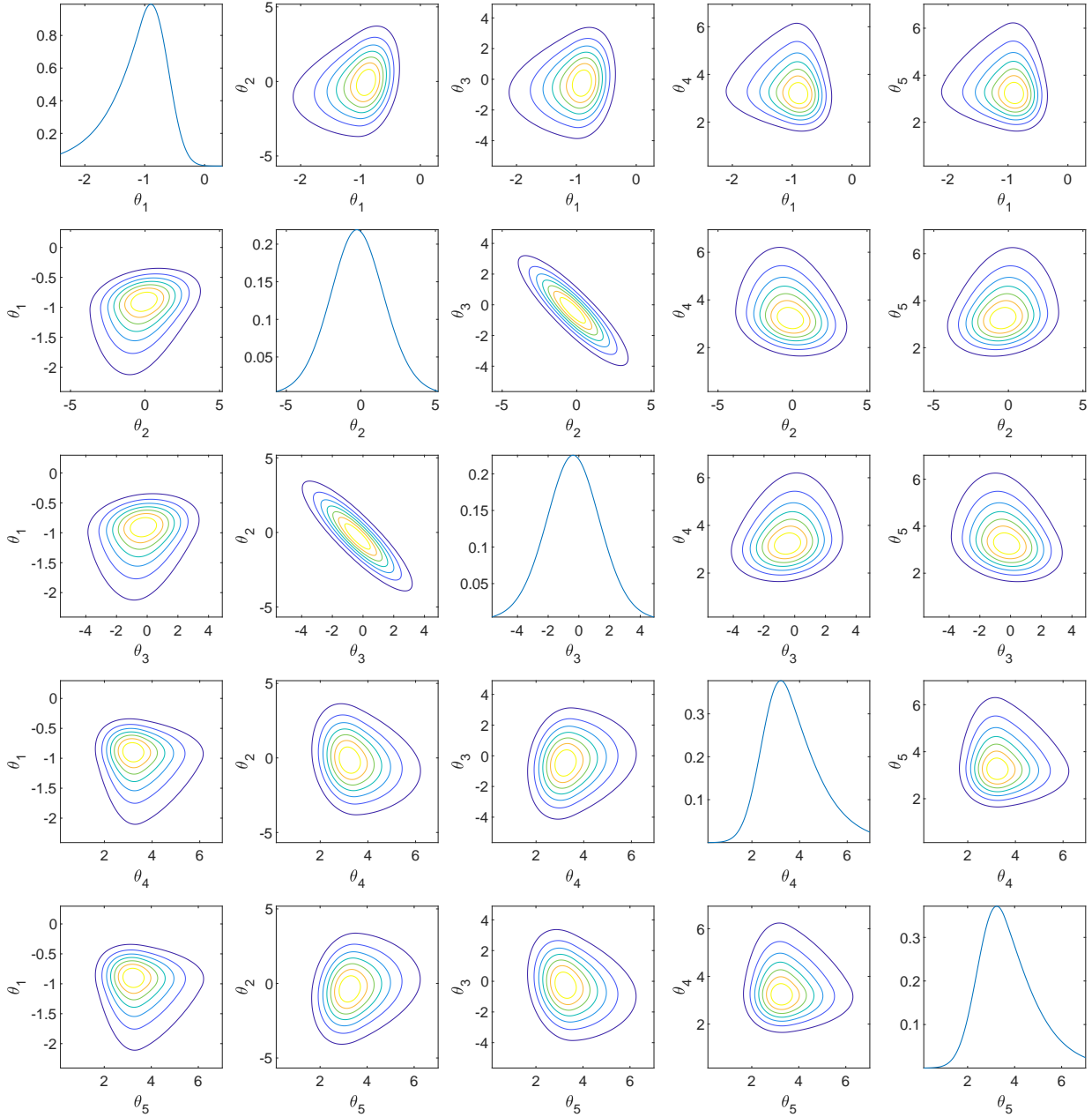}
	\end{center}
	Posterior estimates of $(\theta_1,\ldots,\theta_5)^\top=(\tau_1,\log \chi_1, \log \nu_1, \log \xi, \log \kappa)^\top$
	using the copula model VA with (adjusted) YJ transformations. 
	The marginals $q_{\lambda_i}(\theta_i)$ are on the leading diagonals, and 
	the bivariate marginals---which are bivariate slices of~\eqref{Eq:VA}---are on the off-diagonals. 
	\label{fig:CATX_copVA}
\end{figure}

\begin{figure}[!tbhp]
	\caption{Variational Posterior Densities of Spearman Correlations for Five Western States}
	\begin{center}
		\includegraphics*[width=0.8\textwidth]{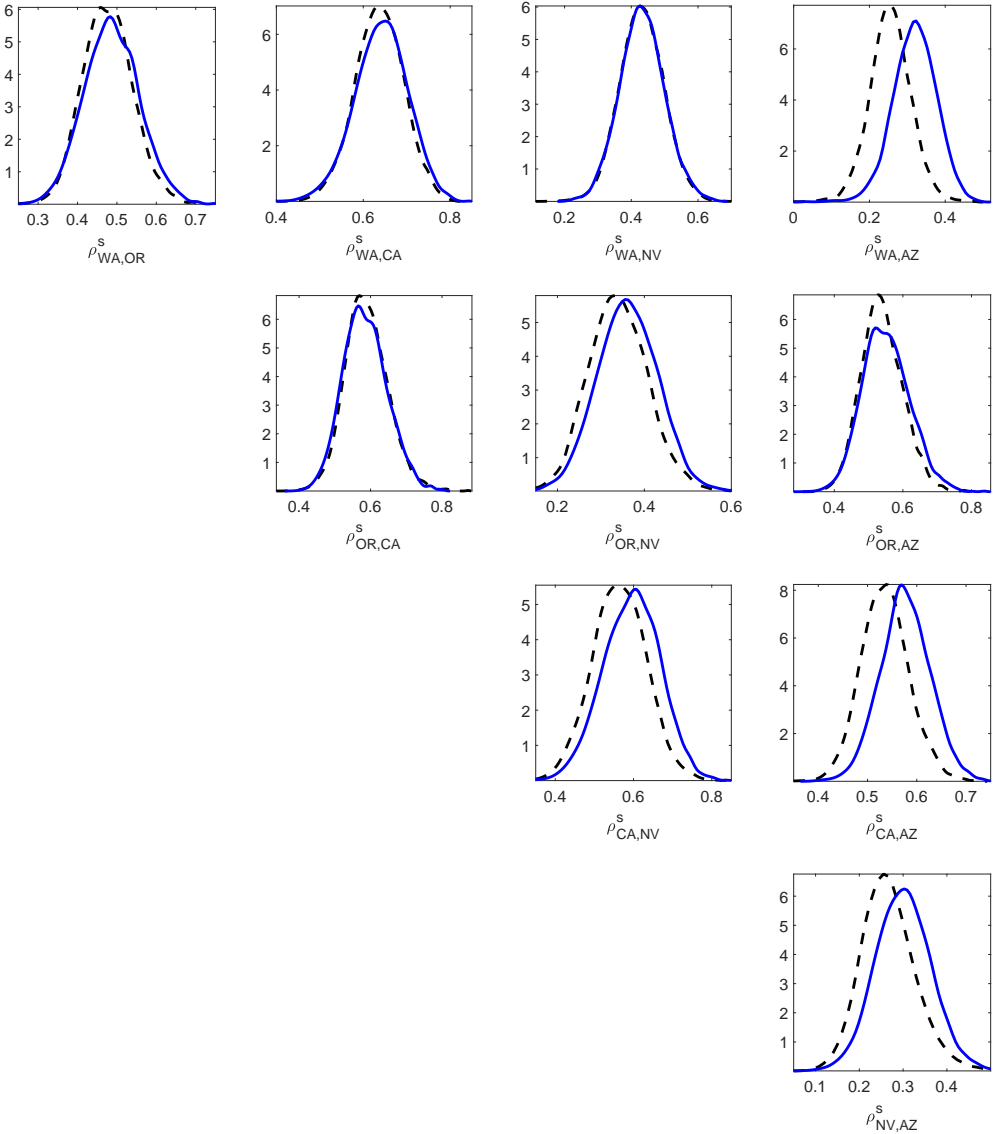}
	\end{center}
Variational posterior densities of the 10 Spearman correlations between the western U.S. states of WA, OR, CA, NV and AZ. The posterior for the
more accurate optimal t-copula is the blue solid line, while that from the mean field t-distribution is the black dashed line. 
	\label{fig:west_spear}
\end{figure}

\FloatBarrier

\newpage
\onehalfspacing
\newpage
\noindent
\setcounter{page}{1}
\begin{center}
	{\bf \Large{Online Appendix for ``Implicit copula variational inference''}}
\end{center}

\vspace{10pt}

\setcounter{figure}{0}
\setcounter{table}{0}
\setcounter{section}{0}
\renewcommand{\thetable}{A\arabic{table}}
\renewcommand{\thefigure}{A\arabic{figure}}

\noindent
This Online Appendix has four parts:

\begin{itemize}
	\item[] {\bf Part~A}: Notational conventions and matrix differentiation rules used.
	\item[] {\bf Part~B}: Additional details on derivations in Appendix~A.
	\item[] {\bf Part~C}: Additional details for the regularized correlation matrix model in Section~4.
	\item[] {\bf Part~D}: Additional empirical results for the U.S. inequality example with 49 states.
	\item[] {\bf Part~E}: List of MATLAB routines provided. 
\end{itemize}
\newpage

\noindent {\bf \large{Part~A: Notational conventions and matrix differentiation rules used}}\\
\ \\
\noindent 
We outline the notational conventions that we adopt in computing
derivatives throughout the paper, which are the same as adopted in
Smith, Loaiza-Maya and Nott~(2020). For a $d$-dimensional vector valued function $g(\bm x)$ of an $n$-dimensional
argument $\bm x$, $\frac{\partial g}{\partial \bm x}$ is the $d\times n$ matrix with element $(i,j)$ $\frac{\partial g_i}{\partial x_j}$.  This means for a scalar $g(\bm x)$, $\frac{\partial g}{\partial \bm x}$ is
a row vector.  When discussing the SGA algorithm we also sometimes write $\nabla_x g(\bm x)=\frac{\partial g}{\partial \bm x}^\top$, which is a column vector.
When the function $g(\bm x)$ or the argument $\bm x$ are matrix valued, then $\frac{\partial g}{\partial \bm x}$ is taken to 
mean $\frac{\partial \text{vec}(g(\bm x))}{\partial \text{vec}(\bm x)}$, where $\text{vec}(A)$ denotes the vectorization of a matrix $A$ obtained by stacking its columns one
underneath another.  If $g(x)$ and $h(x)$ are matrix valued functions, say $g(x)$ takes values which are $d\times r$ and $h(x)$ takes values which are $r\times n$, 
then a matrix valued product rule is
\begin{align*}
	\frac{\partial g(x)h(x)}{\partial x} & = (h(x)^\top\otimes I_d)\frac{\partial g(x)}{\partial x}+(I_n\otimes g(x))\frac{ \partial h(x)}{\partial x}
\end{align*}
where $\otimes$ denotes the Kronecker product and $I_a$ denotes the $a\times a$ identity matrix for a positive integer $a$.  

Some other useful results used repeatedly throughout the derivations below are
$$\text{vec}(ABC)=(C^\top\otimes A)\text{vec}(B),$$
for conformable matrices $A$, $B$ and $C$
the derivative 
\begin{align*}
	\frac{\partial A^{-1}}{\partial A} & = -(A^{-T}\otimes A^{-1}).
\end{align*}
We also write $K_{m,n}$ for the commutation matrix (see, for example, Magnus and Neudecker, 1999).

Last, for scalar function $g(x)$ of scalar-valued argument $x$, we 
sometimes write $g^{\prime}(x)=\frac{d}{d x}g(x)$ and $g^{\prime\prime}(x)
=\frac{d^2}{d x^2}g(x)$ for the first and second derivatives with respect
to $x$ whenever it appears clearer to do so.
\newpage

\noindent {\bf \large{Part~B: Additional details on derivations in Appendix~A}}\\
\ \\

\noindent
Second, we provide further details on the derivation of
$\frac{\partial \bm{\theta}}{\partial \bm{\lambda}}$.
Denote $A = [\bm{d} \ \ B]$ to be the matrix with elements $a_{j,k}$ and $\bm{\varkappa} = \left(\bm{\varkappa}_1^\top,\dots,\bm{\varkappa}_m^\top\right)^\top$.
We need to compute the derivatives  of $\bm{\theta}=h(\bm{\varepsilon},\bm{\lambda})$ with respect to all the elements in $\bm{\lambda} = \left(\bm{\mu}^\top,\bm{\sigma}^\top,\bm{\gamma}^\top,\bm{\tau}^\top\right)^\top $:
\begin{align*}
	\frac{\partial \bm{\theta}}{\partial \bm{\mu}} &= I_m \\
	\frac{\partial \bm{\theta}}{\partial \bm{\sigma}}& = \text{diag}\left(t_{\gamma_1}^{-1}(\psi_1),\dots,t_{\gamma_m}^{-1}(\psi_m)\right)\\
	\frac{\partial \bm{\theta}}{\partial \bm{\gamma}}& = \text{diag}\left(\sigma_1\frac{\partial t_{\gamma_1}^{-1}(\psi_1)}{\partial\bm{\gamma}_1},\dots,\sigma_m\frac{\partial t_{\gamma_m}^{-1}(\psi_m)}{\partial\bm{\gamma}_m}\right)\\
	\frac{\partial \bm{\theta}}{\partial \bm{\tau}} &= \frac{\partial \bm{\theta}}{\partial \bm{\psi}}\frac{\partial \bm{\psi}}{\partial\bm{\varkappa}}\frac{\partial \bm{\varkappa}}{\partial\bm{\tau}}\\
	\frac{\partial \bm{\theta}}{\partial \omegavec} &=\frac{\partial \bm{\theta}}{\partial \bm{\psi}}\frac{\partial \bm{\psi}}{\partial\omegavec} 
\end{align*}
For the last derivative we need to compute the three terms $\frac{\partial \bm{\theta}}{\partial \bm{\psi}}$, $\frac{\partial \bm{\psi}}{\partial\bm{\varkappa}}$, and $\frac{\partial \bm{\varkappa}}{\partial\bm{\tau}}$. The first term can be computed as
$$\frac{\partial \bm{\theta}}{\partial \bm{\psi}} = \text{diag}\left(\sigma_1\frac{\partial t_{\gamma_1}^{-1}(\psi_1)}{\partial\psi_1},\dots,\sigma_m\frac{\partial t_{\gamma_m}^{-1}(\psi_m)}{\partial\psi_m}\right).$$
The third term is given as $\frac{\partial \bm{\varkappa}}{\partial\bm{\tau}} = \text{blockdiag}\left(\frac{\partial \bm{\varkappa}_1}{\partial\bm{\tau}_1},\dots,\frac{\partial \bm{\varkappa}_m}{\partial\bm{\tau}_m}\right)$, with $$\frac{\partial\bm{\varkappa}_j}{\partial\bm{\tau}_j} = \text{diag}\left(\phi(\tau_{j,1})\pi,\dots,\phi(\tau_{j,K-1})\pi,\phi(\tau_{j,K})2\pi\right).$$

For the second term we know that:
\begin{equation}\label{Eq:gradpsi}
	\frac{\partial\bm\psi}{\partial\bm\varkappa} =\sqrt{w}\left[ \left(\bm{z}^\top\otimes I_m\right)\frac{\partial B}{\partial \bm{\varkappa}}+ \text{diag}\left(\bm{\epsilon}\right)\frac{\partial \bm{d}}{\partial \bm{\varkappa}}\right]
\end{equation}
with 
$$\frac{\partial B}{\partial \bm{\varkappa}}=\frac{\partial B}{\partial A^\top}\frac{\partial A^\top}{\partial \bm{\varkappa}}$$
and
$$\frac{\partial \bm{d}}{\partial \bm{\varkappa}}=\frac{\partial \bm{d}}{\partial A^\top}\frac{\partial A^\top}{\partial \bm{\varkappa}}$$
We can write $B = AP_1$ and $\bm{d} = AP_2$ where $P_1 = [\bm{0}_{K\times 1} \ \ I_{K}]^T$ and $P_2 = \left(1,\bm{0}_{1\times K}\right)^\top$.
Then $\frac{\partial B}{\partial A^\top}$ can be easily computed after noting that 
$$\text{vec}\left(B\right)=\left(P_1^\top\otimes I_m\right)\text{vec}\left(A\right)=\left(P_1^\top\otimes I_m\right)K_{K+1,m}\text{vec}\left(A^\top\right)$$
where $K_{K+1,m}$ denotes a commutation matrix. Then
$$\frac{\partial B}{\partial A^\top}=\left(P_1^\top\otimes I_m\right)K_{K+1,m}.$$
The term $\frac{\partial \bm{d}}{\partial A^\top}$ can be computed by noting that
$$\text{vec}\left(\bm{d}\right)=\left(P_2^\top\otimes I_m\right)\text{vec}\left(A\right)=\left(P_2^\top\otimes I_m\right)K_{K+1,m}\text{vec}\left(A^\top\right)$$
so that 
$$\frac{\partial \bm{d}}{\partial A^\top}=\left(P_2^\top\otimes I_m\right)K_{k+1,m}.$$

The elements of the derivative $\frac{\partial A^\top}{\partial \bm{\varkappa}} = \text{blockdiag}\left(\frac{\partial \bm{a}_1}{\partial \bm{\varkappa}_1},\dots,\frac{\partial \bm{a}_m}{\partial \bm{\varkappa}_m}\right)$ can be computed as
\[
\left\{\frac{\partial \bm{a}_i}{\partial \bm{\varkappa}_i}\right\}_{j,l} =
\left\{\begin{array}{cl}
	\cos\left(\varkappa_{i,j}\right)\cos\left(\varkappa_{i,l}\right)\prod_{s\in\{1,\dots,j-1\}\backslash l}\sin\left(\varkappa_{i,s}\right) &\text{If} \ \ \ \ l<j\mbox{ and }  j<K+1\\
	-\prod_{s=1}^{j}\sin\left(\varkappa_{i,s}\right) &\text{If} \ \ \ \ l=j\mbox{ and } j<K+1\\
	\cos\left(\varkappa_{i,l}\right)\prod_{s\in\{1,\dots,j-1\}\backslash l}\sin\left(\varkappa_{i,s}\right) &\text{If} \ \ \ \ l<j\mbox{ and }  j=K+1\\
	0 &\text{Otherwise}\\
\end{array} \right.
\]
Replacing all the terms above into \eqref{Eq:gradpsi} we can show that
$$\frac{\partial\bm\psi}{\partial\bm\varkappa} = \sqrt{w}\left[(\bm{z}^\top P_1^\top)\otimes I_m+ \text{diag}\left(\bm{\epsilon}\right)\left(P_2^\top\otimes I_m\right)\right]K_{K+1,m}\frac{\partial A^\top}{\partial \bm{\varkappa}}.$$

\newpage
\noindent {\bf \large{Part~C: Additional details for the regularized correlation matrix model in Section~4}}\\
\noindent 
This appendix provides further details on the priors and the required gradients and derivatives for the SGA algorithm.

\noindent \textbf{{Priors}}\\
The priors for all the parameters in the model are
$$\tau_{s}~\sim N(0,1),\ \ \ \chi_{s}|\nu_{s}\sim \mathcal{G}^{-1}\left(\frac{1}{2},\frac{1}{\nu_{s}}\right),\ \ \ \xi|\kappa\sim \mathcal{G}^{-1}\left(\frac{1}{2},\frac{1}{\kappa}\right)$$ 
$$ \nu_{1},\dots,\nu_{r(r-1)/2},\kappa\sim \mathcal{G}^{-1}\left(\frac{1}{2},20\right).$$ 

To conduct variational inference we transform all parameters to the real line as follows:
$$\text{(i) }\chi_{s} \ \text{is transformed to } \tilde{\chi}_{s} = \log \chi_{s}; \ \ \ \ \ \ \ \ \ \ \ \ \ \  \text{(ii) }\xi \ \text{is transformed to } \tilde{\xi} = \log \xi;$$
$$\ \text{(iii) }\nu_{s} \ \text{is transformed to } \tilde{\nu}_{s} = \log \nu_{s}; \ \ \ \ \ \ \ \ \ \ \ \ \ \  \ \text{(iv) }\kappa \ \text{is transformed to } \tilde{\kappa} = \log \kappa.$$
After the transformations we obtain the following prior density functions (using
the Jacobians of the change of variables)
\begin{align*}
	\text{(i)}\hspace{0.3cm}&p(\tau_{s}) = \phi_1\left(\tau_{s};0,1\right);\hspace{4cm}\text{(ii)}\ p(\tilde{\chi}_{s}|\nu_{s}) \propto \left(\frac{1}{\nu_{s}}\right)^{0.5}\exp\left(-\frac{1}{2}\tilde{\chi}_{s}-\frac{1}{\nu_{s}\exp(\tilde{\chi}_{s})}\right);\\
	\text{(iii)}\hspace{0.3cm}&p(\tilde{\xi}|\kappa) \propto \left(\frac{1}{\kappa}\right)^{0.5}\exp\left(-\frac{1}{2}\tilde{\xi}-\frac{1}{\kappa\exp(\tilde{\xi})}\right);\hspace{0.1cm}\text{(iv)}\ p(\tilde{\nu}_{s}) \propto \exp\left(-\frac{1}{2}\tilde{\nu}_{s}-\frac{20}{\exp(\tilde{\nu}_{s})}\right);\\
	\text{(v)}\hspace{0.4cm}&p(\tilde{\kappa}) \propto \exp\left(-\frac{1}{2}\tilde{\kappa}-\frac{20}{\exp(\tilde{\kappa})}\right).
\end{align*}
\textbf{{Log-posterior}}\\
\begin{align}
	\log g(\bm{\theta})=&\log p(\bm{y}|\bm{\theta}) +\log p(\bm{\theta})\\
	  \propto& -\frac{N}{2}\log |\Omega|-\frac{1}{2}\sum_{i = 1}^{N}\bm{x}_i^\top\left(\Omega^{-1}-I_r\right)\bm{x}_i+\log p(\bm{\theta})
\end{align}
where the term $\log(\prod_{j=1}^r g_j(y_{i,j}))$ is a constant with respect to $\thetavec$.

\noindent \textbf{{Gradient}}\newline
The model-specific gradient vector for the parameters of the copula model is:
\begin{align*}
	\nabla_{\theta}\log g(\bm{\theta})=  \left(\right.&\nabla_{\tau}\log g(\bm{\theta})^\top,\nabla_{\tilde{\chi}}\log g(\bm{\theta})^\top,\nabla_{\tilde{\xi}}\log g(\bm{\theta})^\top,\nabla_{\tilde{\nu}}\log g(\bm{\theta})^\top,\left.\nabla_{\tilde{\kappa}}\log g(\bm{\theta})^\top\right)^\top.
\end{align*}
The different terms in this gradient can be computed as
\begin{align*}
	\nabla_{\tau}\log g(\bm{\theta})           &  = \left( \sqrt{\xi}\sqrt{\bm{\chi}}\right)\circ\frac{\partial\log g(\bm{\theta})}{\partial {\bm{\eta}}}^\top-\bm{\tau}\\
	\nabla_{\tilde{\chi}}\log g(\bm{\theta})   &  =\frac{1}{2}\bm{\tau}\circ\left( \sqrt{\xi}\sqrt{\bm{\chi}}\right)\circ\frac{\partial\log g(\bm{\theta})}{\partial {\bm{\eta}}}^\top+\nabla_{\tilde{\chi}}\log p(\tilde{\bm{\chi}}|\tilde{\bm{\nu}})\\
	\nabla_{\tilde{\nu}}\log g(\bm{\theta})    &  = \nabla_{\tilde{\nu}}\log p(\tilde{\bm{\chi}}|\tilde{\bm{\nu}}) + \nabla_{\tilde{\nu}}\log p(\tilde{\bm{\nu}})\\
	\nabla_{\tilde{\xi}}\log g(\bm{\theta})    &  = \frac{1}{2}\frac{\partial\log g(\bm{\theta})}{\partial {\bm{\eta}}}\left(\bm{\tau}\circ \sqrt{\xi}\sqrt{\bm{\chi}}\right)-\frac{1}{2}+\frac{1}{\kappa\xi}\\
	\nabla_{\tilde{\kappa}}\log g(\bm{\theta}) &  =-\frac{1}{2}+\frac{1}{\kappa\xi}-\frac{1}{2}+\frac{1}{\kappa}
\end{align*}
where  $\nabla_{\tilde{\chi}}\log p(\tilde{\bm{\chi}}|\tilde{\bm{\nu}}) = \left(-\frac{1}{2}+\frac{1}{\nu_{1}\chi_{1}},\dots,-\frac{1}{2}+\frac{1}{\nu_{r(r-1)/2}\chi_{r(r-1)/2}}\right)^\top$,\\ $\nabla_{\tilde{\nu}}\log p(\tilde{\bm{\chi}}|\tilde{\bm{\nu}}) = \left(-\frac{1}{2}+\frac{1}{\nu_{s}\chi_{s}},\dots,-\frac{1}{2}+\frac{1}{\nu_{r(r-1)/2}\chi_{r(r-1)}}\right)^\top$,  $\nabla_{\tilde{\nu}}\log p(\tilde{\bm{\nu}}) = \left(-\frac{1}{2}+\frac{1}{\nu_{r(r-1)/2}},\dots,-\frac{1}{2}+\frac{1}{\nu_{r(r-1)/2}}\right)^\top$.

$$\frac{\partial\log g(\bm{\theta})}{\partial {\bm{\eta}}} = \left[-\frac{N}{2}\text{vec}(\Omega^{-1})^\top\frac{\partial\Omega}{\partial L}-\frac{1}{2}\left(\sum_{i=1}^{N}\bm{x}_i^\top\otimes\bm{x}_i^\top\right)\frac{\partial\Omega^{-1}}{\partial\Omega}\frac{\partial \Omega}{\partial L}\right]\frac{\partial L}{\partial \bm{\vartheta}}\frac{\partial\bm{\vartheta}}{\partial\bm{\eta}}$$
$$\frac{\partial\Omega}{\partial L} = \left(I_{r^2}+K_{r,r}\right)(L\otimes I_r)$$
$$\frac{\partial\Omega^{-1}}{\partial \Omega} = -\Omega^{-1}\otimes \Omega^{-1}$$
$$\frac{\partial \bm{\vartheta}}{\partial {\bm{\eta}}} = \text{blockdiag}\left(\frac{\partial \bm{\vartheta}_2 }{\partial\bm{\eta}_2},\dots,\frac{\partial \bm{\vartheta}_r }{\partial\bm{\eta}_r}\right)$$
$$\frac{\partial \bm{\vartheta}_i}{\partial\bm{\eta}_i} = \text{diag}\left[\phi(\eta_{i,1})\pi,\dots,\phi\left(\eta_{i,i-1}+\Phi^{-1}\left(\frac{1}{4}\right)\right)2\pi\right]$$

Finally, $\frac{\partial L}{\partial \bm{\vartheta}}$ is a matrix with elements $\left\{\frac{\partial L}{\partial \bm{\vartheta}}\right\}_{(j-1)r+i,(i-1)(r-1)+k} = \frac{\partial l_{i,j}}{\partial \vartheta_{i,k}}$, where
\[
\frac{\partial l_{i,j}}{\partial \vartheta_{i,k}} =
\left\{\begin{array}{cl}
	\cos\left(\vartheta_{i,i-1}\right)\prod_{s=1}^{i-2}\sin\left(\vartheta_{i,s}\right) &\text{If} \ \ \ \ i>1\mbox{ and }  j=i\mbox{ and }  k = i-1\\	\sin\left(\vartheta_{i,i-1}\right)\frac{\cos\left(\vartheta_{i,k}\right)}{\sin\left(\vartheta_{i,k}\right)}\prod_{s=1}^{i-2}\sin\left(\vartheta_{i,s}\right) &\text{If} \ \ \ \ i>1\mbox{ and }  j=i\mbox{ and } k < i-1\\
	-\prod_{s=1}^{j}\sin\left(\vartheta_{i,s}\right)&\text{If} \ \ \ \ i>1\mbox{ and }  j<i\mbox{ and } k = j\\
	\cos\left(\vartheta_{i,j}\right)\frac{\cos(\vartheta_{i,k})}{\sin(\vartheta_{i,k})}\prod_{s=1}^{j-1}\sin\left(\vartheta_{i,s}\right)&\text{If} \ \ \ \ i>1\mbox{ and }  j<i\mbox{ and } k < j\\
	0 &\text{otherwise}.\\
\end{array} \right.
\]
If not specified, all remaining elements of $\frac{\partial L}{\partial \bm{\vartheta}}$ are zero.

\newpage
\noindent {\bf \large{Part~D: Additional empirical results for the U.S. inequality example with 49 states}}\\
 
\begin{figure}[!htbp]
	\caption{Marginal Data Density Estimates for Nine Most Populous U.S. States}\label{fig:afig1}
	\begin{center}
		\includegraphics*[width=\textwidth]{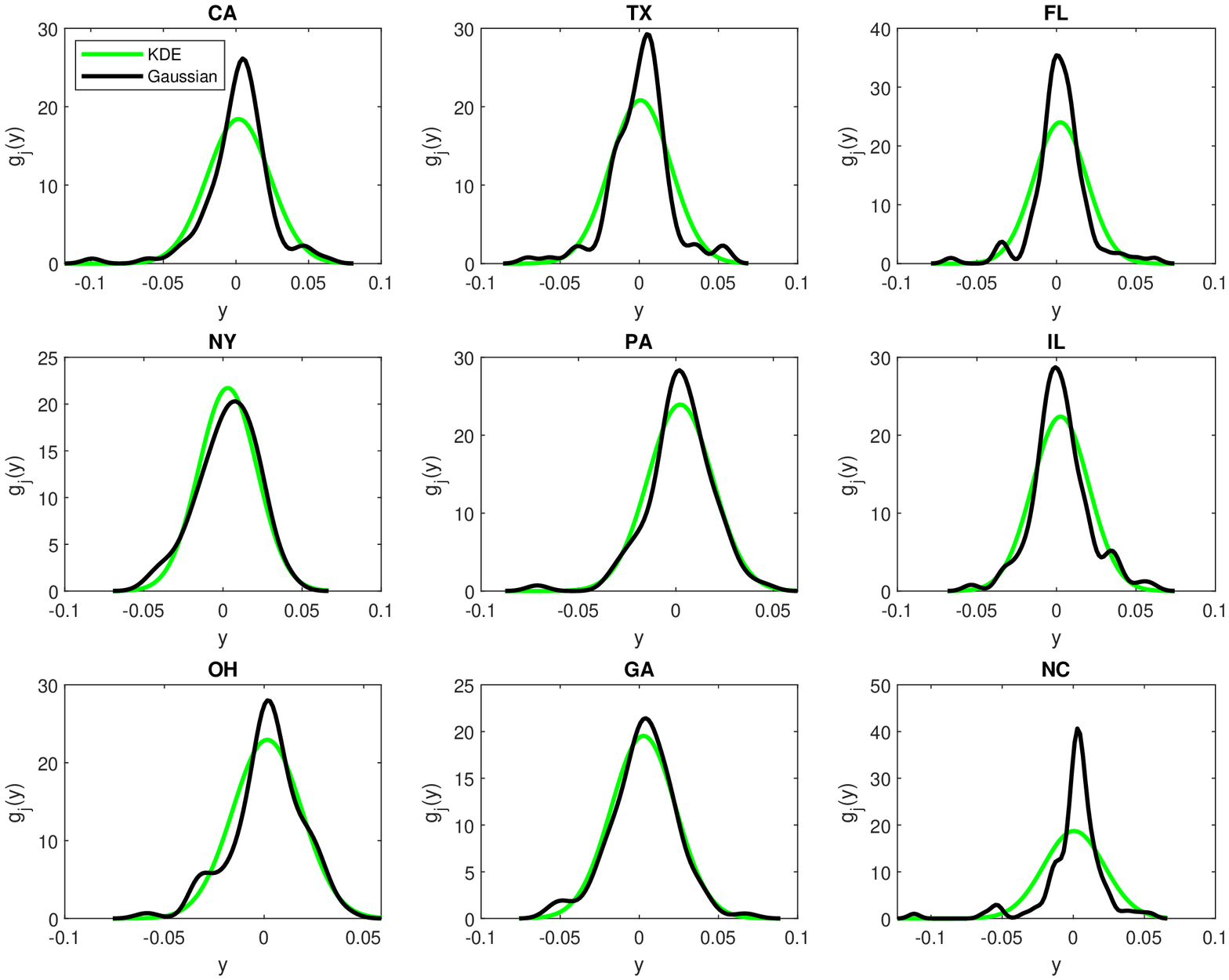}
	\end{center}
	Marginal density estimates of $g_j(y)=\frac{d}{dy}G_j(y)$ for the nine most populous 
	U.S. states. In each panel the KDE (black line) is used to compute the copula data $u_{t,j}=G_j(y_{t,j})$,
	while the fitted Gaussian (green line) is plotted for comparison. 
\end{figure}

\noindent
Here we include some additional empirical results for the U.S. income inequality example. Figure~\ref{fig:afig1} provides the kernel density
estimates of the
marginal
densities $g_j$ for the nine most popular U.S. states. These are far from
Gaussian, so that a Gaussian copula model that accounts for this is more appropriate than fitting a multivariate normal distribution. The KDE's for the remaining 40 states (unreported) are also far from Gaussian. 

\begin{figure}[!htbp]
	\caption{Variational Posterior Mean of Spearman Correlations in U.S. Inequality Example}\label{fig:afig2}
	\begin{center}
		\includegraphics*[width=\textwidth]{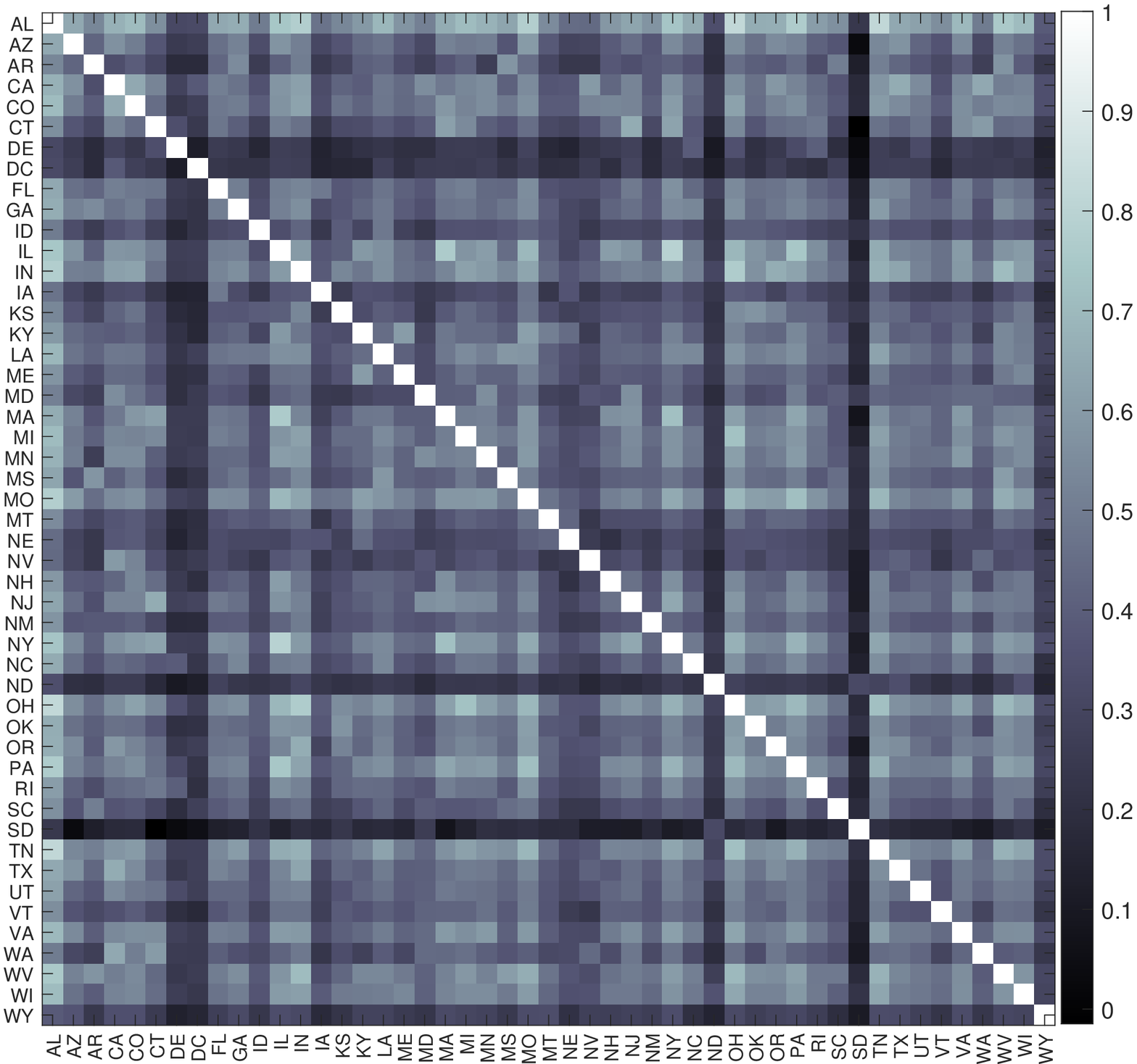}
	\end{center}
	Posterior mean estimates of Spearman correlation matrix $\Omega^s$ for all 49
	U.S. states in our data. These estimates were obtained by simulating draws from $q_{\lambda^\star}(\thetavec)$, evaluating the resulting draws of $\Omega^s$
	and computing their mean. 
\end{figure}

Figure~\ref{fig:afig2}	reports the posterior mean estimates of Spearman correlation matrix $\Omega^s$ for all 49
U.S. states in our data. These estimates were obtained by simulating draws from $q_{\lambda^\star}(\thetavec)$, evaluating the resulting draws of $\Omega^s$
and computing their mean. There are no negative Spearman correlations, and 
the highest dependencies tend to be between geographically adjacent or close states. 
For example, IL is positively dependent with NY, PA and MA. 

\begin{landscape}
\begin{figure}[!htbp]
	\caption{Evidence Lower Bound for VAs in U.S. Inequality Example}\label{fig:afig3}
	\begin{center}
		\includegraphics*[scale=0.8]{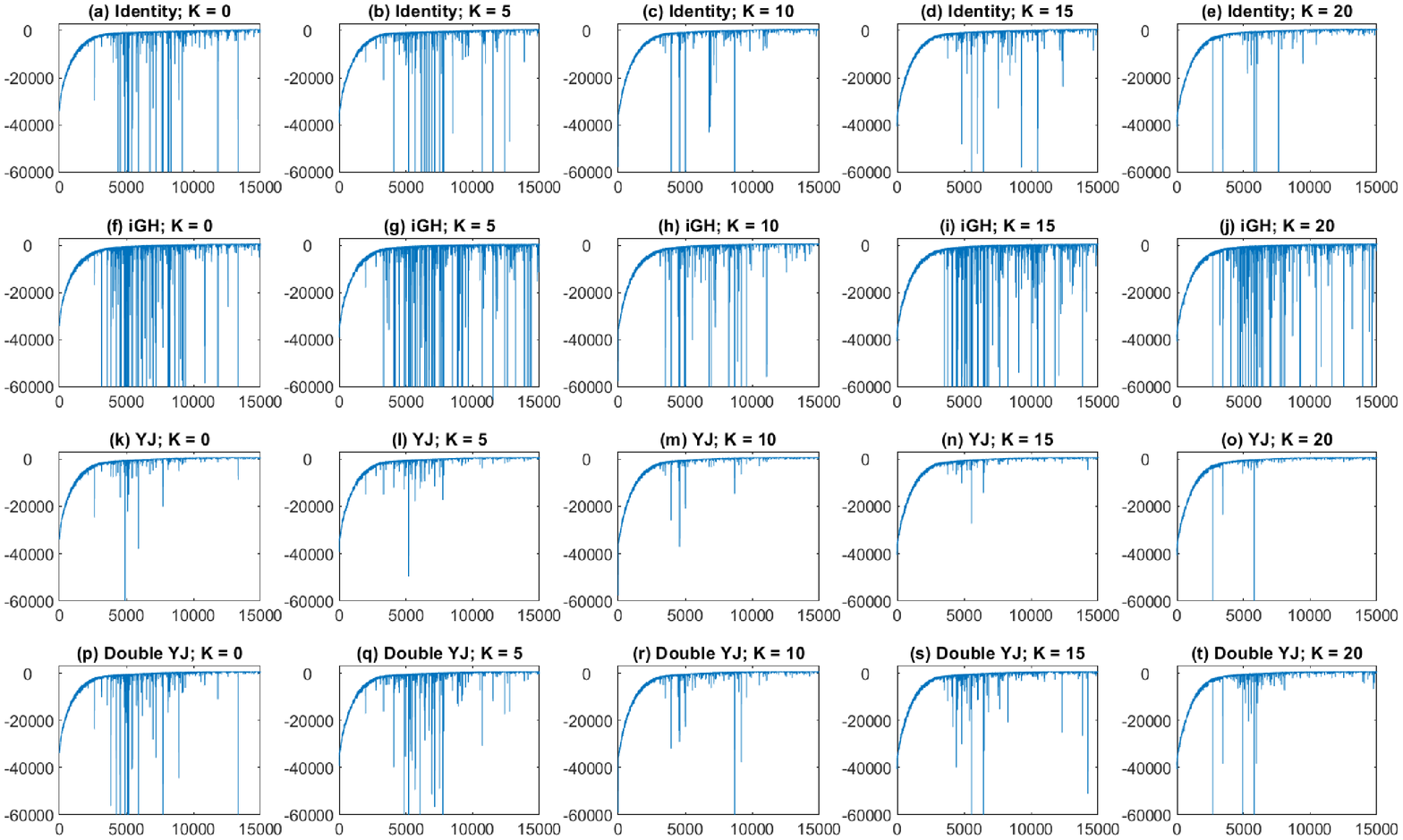}
	\end{center}
	Rows correspond to different transformations and columns correspond to different number of factors in factor covariance structure.
\end{figure}
\end{landscape}

Figure~\ref{fig:afig3} plots the approximate ELBO values against 
step number for the
twenty VAs in Table~2 of the manuscript. The lower bound value is approximate because in the SGA algorithm with 
the re-parameterization gradient,
at each step we have a single
draw from $f_\varepsilon$ and evaluate
\begin{eqnarray*}
{\cal L}(q_\lambda) &= &E_{f_\varepsilon}\left[
\log g(h(\varepsilonvec,\lambdavec)) - \log q_\lambda (h(\varepsilonvec,\lambdavec))
\right]\\
&\approx & 
\log g(h(\varepsilonvec,\lambdavec)) - \log q_\lambda (h(\varepsilonvec,\lambdavec))\,,\mbox{ for } \varepsilonvec \sim f_\varepsilon
\end{eqnarray*}
This makes the trace plots noisy, which is also the case for those
produced in~Ong, Nott and Smith~(2018), and~Smith, Loaiza-Maya and~Nott~(2020). Generating multiple
draws from $f_\varepsilon$ and averaging produces a less noisy approximation 
of the ELBO function, but slows the SGA algorithm dramatically and is
typically not done when using the re-parameterization gradient. This is also why we report the median ELBO values over the last 500 steps (i.e. $\overline{\mbox{LB}}$) in Table~2.
Notice that
the SGA algorithm appears to converge by this metric in around 5000 - 10,000
steps for all 20 VAs considered. Each panel gives the trace for a different
VA, and the arrangement of the panels matches that of the reported 
values of $\overline{\mbox{LB}}$ in Table~2 of the manuscript.

\newpage
\noindent {\bf \large{Part~E: List of MATLAB routines provided}}

\noindent 
As part of the Supplementary Material, we provide the MATLAB code and data 
employed in this study. Details on how to replicate our results using this
code are given in README files. Below we also provide a list of the main MATLAB routines employed. Note that these can be used to 
implement our
method with other statistical models.

\begin{itemize}
		\item[1.] High level functions used to implement our variational inference approach include:
	\begin{itemize}
		\item[] \textbf{VB\_step}: updates the variational parameter according to $\bm\lambda^{(s+1)} = \bm\lambda^{(s)}+\bm{\delta}^{(s)}\circ\widehat{\nabla_\lambda\mathcal{L}(q_\lambda)}$, for the Gaussian copula VA 
		\item[] \textbf{VB\_step\_tcop}: updates the variational parameter according to $\bm\lambda^{(s+1)} = \bm\lambda^{(s)}+\bm{\delta}^{(s)}\circ\widehat{\nabla_\lambda\mathcal{L}(q_\lambda)}$, for the t-copula VA
		\item[] \textbf{VBtransf.m}: Implements variational inference using the proposed copula approximation
	\end{itemize}
	\item[2.] Functions used to evaluate the re-parameterization gradient efficiently:
		\begin{itemize}
		\setlength{\itemindent}{-.25in}
		\item[] \textbf{dxT\_dkappa}: compute term $\frac{\partial A^\top}{\partial\bm\varkappa}$
		\item[] \textbf{dBd\_dkappa.m}: compute the terms $\frac{\partial B}{\partial \bm{\varkappa}}$ and $\frac{\partial \bm d}{\partial \bm{\varkappa}}$ 
		\item[] \textbf{dthetadkappa.m}: compute the term $\frac{\partial \bm\theta}{\partial \bm{\varkappa}} = \frac{\partial \bm\theta}{\partial \bm{\psi}} \frac{\partial \bm\psi}{\partial \bm{\varkappa}}$
		\item[] \textbf{dkappa\_dtkappa.m}: compute the term $\frac{\partial \bm \varkappa}{\partial \bm{\tau}}$
		\item[] \textbf{dwdnu.m}: compute expression $\frac{\partial}{\partial \nu}F_W^{-1}(u;\nu)$ where $F_W$ is the t-distribution cdf
		\item[] \textbf{dphidnu.m}: compute the term $\frac{\partial \bm\psi}{\partial \nu}$ for t-copula approximation
	    \item[] \textbf{dthetadnu.m}: compute the term $\frac{\partial \bm\theta}{\partial \nu} = \frac{\partial \bm\theta}{\partial \bm{\psi}}\frac{\partial \bm\psi}{\partial \nu}$ for t-copula approximation
	    \item[] \textbf{dtheta\_dtau}: compute the term $\frac{\partial\bm\theta}{\partial\bm\gamma}$
	    \item[] \textbf{grad\_theta\_logq.m}: compute term $\nabla_\theta \log q_\lambda(\bm{\theta})$ for the Gaussian copula VA
	    \item[] \textbf{grad\_theta\_logq\_tcop.m}: compute term $\nabla_\theta \log q_\lambda(\bm{\theta})$ for the t-copula VA
	    \item[] \textbf{gradient\_compute.m}: compute $\widehat{\nabla_\lambda\mathcal{L}(q_\lambda)} =\left.\frac{dh(\varepsilonvec,\lambdavec)}{d\lambdavec}\right.^\top \left( \nabla_\theta
	    \log g(\thetavec) - \nabla_\theta \log q_\lambda (\thetavec)\right)$ for the Gaussian copula VA
	    \item[] \textbf{gradient\_compute\_tcop.m}: compute $\widehat{\nabla_\lambda\mathcal{L}(q_\lambda)} = \left.\frac{dh(\varepsilonvec,\lambdavec)}{d\lambdavec}\right.^\top \left( \nabla_\theta
\log g(\thetavec) - \nabla_\theta \log q_\lambda (\thetavec)\right)$ for the t-copula VA
	\end{itemize}
	\item[3.] Low-level functions associated with the iGH and YJ transformations:
\begin{itemize}
	\setlength{\itemindent}{-.25in}
	\item[] \textbf{gh.m}: compute the G\&H transformation
	\item[] \textbf{igh.m}: compute the inverse G\&H transformation
	\item[] \textbf{dgh\_g.m dgh\_h.m}: compute the derivatives of the G\&H transformation with respect to g and h, respectively
	\item[] \textbf{dgh.m}: compute the first derivative of the G\&H transformation with respect to its argument
	\item[] \textbf{dgih.m}: compute the first derivative of the inverse G\&H transformation with respect to its argument
	\item[] \textbf{ddgh.m}: compute the second derivative of the G\&H transformation with respect to its argument
	\item[] \textbf{ddigh.m}: compute the second derivative of the inverse G\&H transformation with respect to its argument
	\item[] \textbf{YJ.m}: compute the YJ transformation
	\item[] \textbf{iYJ.m}: compute the inverse YJ transformation
	\item[] \textbf{diYJ\_deta.m}: compute the derivatives of the YJ transformation with respect to $\gamma$ parameter
	\item[] \textbf{dYJ.m}: compute the first derivative of the YJ transformation with respect to its argument
	\item[] \textbf{diYJ.m}: compute the first derivative of the inverse YJ transformation with respect to its argument
	\item[] \textbf{ddYJ.m}: compute the second derivative of the YJ transformation with respect to its argument	
\end{itemize}
\end{itemize} 
\end{document}